\documentclass[aps,pra,reprint,superscriptaddress,twocolumn,showkeys,amsmath,amssymb,longbibliography]{revtex4-2}
\usepackage[english]{babel}
\usepackage{amsmath,amssymb,bbm,mathrsfs,bm,braket,color,graphicx,comment,amsfonts,dsfont}
\usepackage[colorlinks,citecolor=blue,urlcolor=blue]{hyperref}
\usepackage[mathscr]{euscript}
\usepackage[normalem]{ulem}
\usepackage{comment}

\begin{document}
\title{Network model for magnetic higher-order topological phases}

\author{Hui Liu}
\affiliation{IFW Dresden and W{\"u}rzburg-Dresden Cluster of Excellence ct.qmat, Helmholtzstrasse 20, 01069 Dresden, Germany}
\affiliation{Department of Physics, Stockholm University, AlbaNova University Center, 106 91 Stockholm, Sweden}

\author{Ali G. Moghaddam}
\affiliation{Department of Physics, Institute for Advanced Studies in Basic Sciences (IASBS), Zanjan 45137-66731, Iran}
\affiliation{Research Center for Basic Sciences $\&$ Modern Technologies (RBST), Institute for Advanced Studies in Basic Science (IASBS), Zanjan 45137-66731, Iran}

\author{Daniel Varjas}
\affiliation{IFW Dresden and W{\"u}rzburg-Dresden Cluster of Excellence ct.qmat, Helmholtzstrasse 20, 01069 Dresden, Germany}
\affiliation{Department of Physics, Stockholm University, AlbaNova University Center, 106 91 Stockholm, Sweden}
\affiliation{Max Planck Institute for the Physics of Complex Systems, Nöthnitzer Strasse 38, 01187 Dresden, Germany}
\affiliation{Department of Theoretical Physics, Institute of Physics, Budapest University of Technology and Economics, Műegyetem rkp. 3., H-1111 Budapest, Hungary}

\author{Ion Cosma Fulga}
\affiliation{IFW Dresden and W{\"u}rzburg-Dresden Cluster of Excellence ct.qmat, Helmholtzstrasse 20, 01069 Dresden, Germany}

\begin{abstract}
We propose a network-model realization of magnetic higher-order topological phases (HOTPs) in the presence of the combined space-time symmetry $C_4\mathcal{T}$ -- the product of a fourfold rotation and time-reversal symmetry. 
We show that the system possesses two types of HOTPs. 
The first type, analogous to Floquet topology, generates a total of $8$ corner modes at $0$ or $\pi$ eigenphase, while the second type, hidden behind a weak topological phase, yields a unique phase with $8$ corner modes at $\pm\pi/2$ eigenphase (after gapping out the counterpropagating edge states), arising from the product of particle-hole and phase-rotation symmetry.
By using a bulk $\mathbb{Z}_4$ topological index ($Q$), we found both HOTPs have $Q=2$, whereas $Q=0$ for the trivial and the conventional weak topological phase. 
Together with a $\mathbb{Z}_2$ topological index associated with the reflection matrix, we are able to fully distinguish all phases.
Our work motivates further studies on magnetic topological phases and symmetry protected $2\pi/n$ boundary modes, as well as suggests that such phases may find their experimental realization in coupled-ring-resonator networks.

\end{abstract}
\maketitle

\section{Introduction}
\label{sec:intro}

Topological phases of matter are generically diagnosed by certain invariants revealing topological characteristics of bulk states.
According to bulk-edge correspondence, nontrivial band topologies are associated with gapless boundary modes~\cite{Kane2010RMP, Zhang2011RMP}. 
Most of the topological states rely on the presence of certain symmetries, an aspect which has led to the tenfold classification of symmetry-protected topological phases by considering fundamental symmetries, i.e., time-reversal, particle-hole, and chiral symmetry~\cite{Altland1997, Schnyder2009, Ryu2010, Chiu2016}. 
Recently, a new category of topological phases has been found, in which topologically protected boundary states of a $d$-dimensional material have a dimension less than $(d-1)$~\cite{Benalcazar2017, Benalcazar2017prb, Langbehn2017, Song2017}. 
These phases are dubbed as higher-order topological phases (HOTPs) and can be realized in crystalline and quasi-crystalline structures, where point group symmetries protect corner or hinge modes  \cite{Schindler2018natphys, Schindler2018sciadv, Geier2018, Khalaf2018, Trifunovic2019, Hughes2018, Ortix2018, You2018, Roy2019, Araki2019, Hatsugai2019, Tiwari2020, Ezawa2018kagome, Ezawa2018phosphorene, loss2018, Wang2018prl, Yan2018prl,xie2021higher}.

The interplay of crystalline symmetries and fundamental symmetries is of particular interest, as it can result in new phases of matter. 
A prime illustration is observed in magnetic topological materials~\cite{Tokura2019, Elcoro2021, Bernevig2022}, where although local moments explicitly break time reversal symmetry, the antiunitarity of symmetry operations is still preserved through combining with crystalline symmetries. 
An existing example is the antiferromagnetic topological insulator~\cite{ati, Otrokov2019}, in which the combination of time-reversal symmetry and translational or inversion symmetry leads to a quantized magnetoelectric effect. 
Besides that, very recently, a new direction has emerged in the fundamental magnetism, the so-called altermagnets~\cite{altermagnetism1,altermagnetism2,Guo2023}, where the product of rotation and time-reversal symmetry plays a central role. 

Network models consisting of scattering arrays provide a valuable tool for simulating topological phases of matter. 
It was first introduced in the seminal work by Chalker and Coddington to study the robustness of topological states in quantum Hall systems~\cite{chalker1988, kramer2005review}, and based on its advantage in numerics, has been subsequently extended to a variety of topological systems, both in the context of tenfold classification~\cite{chalker1999super, chalker2001super, chalker2002qshe, murdy2007qshe, murdy2014TI, fulga2012thermal} and point group symmetry protected HOTPs~\cite{Liu2020-hoti}. 
Network models deal with amplitudes of propagating modes and are described by unitary operators, which relate them to Floquet topology~\cite{ho1996, zirnbauer1999, potter2020}, the topology of discrete-time evolution operators. 
However, to our best knowledge, there is an important ingredient in general unique for network models, dubbed as the phase rotation symmetry~\cite{Phase_rotation1, Phase_rotation2}. 
It links an eigenphase $\varepsilon$ at momentum $\mathbf{k}$ to an eigenphase $\varepsilon+\phi$ at the same momentum, without changing any other quantities, which could also lead to new types of topologically protected phases.

In this work, we study magnetic topological phases in a network model, protected by the combination of fourfold rotation symmetry $C_4$ and time-reversal symmetry $\mathcal{T}$.
This model is a spinful version of the Chalker-Coddington network for superconductors (class D in tenfold symmetry classification~\cite{Altland1997}). 
By imposing the $C_4{\mathcal T}$ symmetry constraint to the network, we find four different phases: a HOTP, a hidden-HOTP weak topological phase (WTP), a conventional WTP, and a trivial phase. 
In the HOTP, corner modes similar to those in Floquet topological systems stay at $0$ or $\pi$ eigenphase. 
However, in the hidden-HOTP WTP, dimerization gapping out the counterpropagating edge states will lead to a new type of corner modes, which have eigenphases $\pm\pi/2$, as a consequence of the product of particle-hole symmetry and twofold phase rotation symmetry. 
We use the pffafian invariant, a $\mathbb{Z}_4$ index, recently introduced in Ref.~\cite{pfaffian} to distinguish different phases of the ${\cal C}_4{\cal T}$-symmetric network. This index takes the value $Q=2$ for the HOTP and the hidden-HOTP WTP, and $Q=0$ for the conventional WTP and the trivial phase. 
A further classification can be applied when both the bulk and the edges of the finite-sized network are gapped by the dimerization (which breaks translational symmetry), where the determinant of the corner reflection matrix provides another topological invariant taking the values $\nu=\pm 1$ for the HOTP and the hidden HOTP, respectively.

The rest of this work is organized as follows. In Sec.~\ref{sec:model} we introduce the network model and its symmetries. 
The formation of a HOTP can be intuitively understood from a special point in parameter space, the so-called decoupled limit, which we analyze in Sec.~\ref{sec:HOTP}. 
We determine the full phase diagram in Sec.~\ref{sec:phase_diagram} and confirm the topological nature of the different phases by computing their invariants in Sec.~\ref{sec:top_inv}. 
The hidden higher-order topological phase is discussed in Sec.~\ref{sec:hidden}, before concluding in Sec.~\ref{sec:conc}.

\section{Model construction and symmetry}
\label{sec:model}

\begin{figure}
\centering
\includegraphics[width=\linewidth]{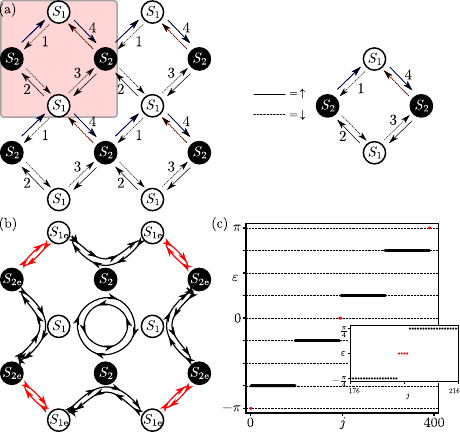}
\caption{
On the top left we show a sketch of a finite-sized network model, whose unit cell (the shaded area) is shown on the top right. 
The arrows represent Majorana modes with wavefunctions $\psi_{i\sigma}$ ($i=1$, $2$, $3$, $4$, $\sigma=\uparrow$, $\downarrow$). 
Their intersections represent scattering nodes $S_1$ and $S_2$. 
The bottom left and right are an example of a configuration producing a higher-order topological phase and its eigenphase spectrum with a system size $7\times 7$ unit cells, respectively. 
\label{fig:model}
}
\end{figure}

The network model is a square lattice made of directed links and scattering nodes (see Fig.~\ref{fig:model}(a)).
Each link corresponds to a propagating Majorana mode (shown by solid and dashed arrows), as we are interested in describing a particle-hole symmetric system.
Each of the two inequivalent nodes connects four incoming modes to four outgoing modes and can be described by a unitary $4\times 4$ scattering matrix ($S_1$ and $S_2$, respectively).
A unit cell of the network consists of eight such propagating modes, each of which represents a component of the full network wavefunction, 
$\Psi=(\psi_{1\uparrow},\psi_{2\downarrow},\psi_{3\uparrow},\psi_{4\downarrow},\psi_{1\downarrow},\psi_{2\uparrow},\psi_{3\downarrow},\psi_{4\uparrow})^T$.

Considering an infinite, translationally-invariant network, we can describe the evolution of the network wavefunction upon successive scattering events in momentum space. 
The steady states of the network are those which retain their structure upon multiple scattering, being described by the eigenvalue equation 
\begin{eqnarray}
\mathcal{S}(\mathbf{k})\Psi(\mathbf{k})=e^{i\varepsilon(\mathbf{k})}\Psi(\mathbf{k}),
\end{eqnarray}  
where $\mathbf{k}=(k_x, k_y)$ are the two momenta, $\Psi(\mathbf{k})$ is the network wavefunction in the basis written above, $\varepsilon(\mathbf{k})$ is its associated eigenphase, and $\mathcal{S}(\mathbf{k})$ is the so-called Ho-Chalker operator, a unitary matrix encoding intracell and intercell scattering events as
\begin{eqnarray}
\mathcal{S}(\mathbf{k})=\begin{pmatrix}
0&p_xS_2p_x^\dagger\\
p_y^\dagger S_1p_y&0
\end{pmatrix},
\end{eqnarray}
with $p_x=\text{diag}[1,1,e^{ik_x},e^{ik_x}]$ indicating scatterings along the $x$-direction and $p_y=\text{diag}[e^{ik_y},1,1,e^{ik_y}]$ representing scatterings along the $y$-direction.

The presence of symmetries further constrains the node scattering matrices, $S_1$ and $S_2$. 
Since we are interested in networks with $C_4\mathcal{T}$ symmetry, we begin by considering fourfold rotations around the center of the unit cell, leading to a $C_4$ symmetry operator of the form
\begin{eqnarray}
\mathcal{U}_{C_4}=\begin{pmatrix}0&1\\-1&0\end{pmatrix}\otimes \tilde{c}_4,~\tilde{c}_4=\begin{pmatrix}
0&1&0&0\\
0&0&-1&0\\
0&0&0&1\\
1&0&0&0
\end{pmatrix}.\label{eq:c4symmetry}
\end{eqnarray}
Within the unit cell, it obeys $\mathcal{U}_{C_4}^4=-\mathbb{I}_{8\times 8}$ ($\mathbb{I}$ denotes the identity matrix) and rotates network wavefunctions as $\psi_{1\sigma}\Rightarrow\psi_{2\sigma}\Rightarrow\psi_{3\sigma}\Rightarrow\psi_{4\sigma}\Rightarrow\psi_{1\sigma}$ ($\sigma=\uparrow, \downarrow$ encodes the spin degree of freedom).

Time-reversal symmetry, on the other hand, involves a local spin flipping ($\psi_{i\uparrow}\rightarrow\psi_{i\downarrow}$, $\psi_{i\downarrow}\rightarrow -\psi_{i\uparrow}$) and reverses the propagation direction, the unitary part of which can be written as 
\begin{eqnarray}
\mathcal{U}_{\mathcal{T}}=-i\tau_y\otimes \mathbb{I}_{4\times 4},
\end{eqnarray}
with $\mathcal{K}$ the complex conjugation and $\tau_i$ ($i=x$, $y$, $z$) Pauli matrices.
Combining both symmetry operators leads to a matrix representation for the unitary rotation part of the $C_4\mathcal{T}$ symmetry,
\begin{eqnarray}
\mathcal{U}_{C_4\mathcal{T}}=\mathbb{I}_{2\times 2}\otimes\tilde{c}_4.
\end{eqnarray}
For the Ho-Chalker (evolution) operator, it implies
\begin{eqnarray}
\mathcal{U}_{C_4\mathcal{T}}\mathcal{S}^{*}(k_x,k_y)\mathcal{U}_{C_4\mathcal{T}}^\dagger=S^{\dagger}(k_y,-k_x).\label{eq:sym_hc}
\end{eqnarray}

To obtain the explicit form of $S_1$ and $S_2$, we first focus on the $C_4\mathcal{T}$ symmetry constraint at high symmetry point $\mathbf{k}=(0, 0)$. 
It leads to
\begin{eqnarray}
\tilde{c}_4 S_2 \tilde{c}_4^T=S_1^T,~\tilde{c}_4 S_1 \tilde{c}_4^T=S_2^T. \label{eq:sym_node}
\end{eqnarray}
This reduces the needed parameter space and provides
\begin{eqnarray}
(\tilde{c}_4)^2 S_i (\tilde{c}_4^T)^2 = S_i,~i=1,~2.\label{eq:sym_node1}
\end{eqnarray} 
We then only need to focus on $S_1$.
In the presence of particle-hole symmetry, $S_1$ is real and can be written as a polar decomposition~\cite{Beenakker_review},
\begin{eqnarray}
S_1=
\begin{pmatrix}o_1&0\\0&o_2\end{pmatrix}
\begin{pmatrix}
c_1&s_1\\s_1&-c_1
\end{pmatrix}
\begin{pmatrix}o_3&0\\0&o_4\end{pmatrix}
,\label{eq:polar_form}
\end{eqnarray}
where $c_i=\cos(\text{diag}[\alpha_{i},\beta_{i}])$, $s_i=\sin(\text{diag}[\alpha_{i},\beta_{i}])$, and $o_i=e^{i\theta_{i}\tau_y}$.
Substituting it in Eq.~\eqref{eq:sym_node1} leads to
\begin{eqnarray}
S_1=
(o_1\otimes\tau_0)
\begin{pmatrix}
c_1&s_1\\s_1&-c_1
\end{pmatrix}
(o_2\otimes\tau_z)
.
\label{eq:s1}
\end{eqnarray}
Further substituting Eqs.~\eqref{eq:sym_node} and \eqref{eq:s1} into Eq.~\eqref{eq:sym_hc}, we find that $S_{1,2}$ indeed respect $C_4\mathcal{T}$ symmetry with a parameter space $(\theta_1,\theta_2,\alpha,\beta)$.

Since both $S_1$ and $S_2$ are real matrices, the corresponding Ho-Chalker operator is also protected by particle-hole symmetry as
\begin{eqnarray}
{\cal S}^{\ast}(-\mathbf{k})={\cal S}(\mathbf{k}),
\end{eqnarray}
which means for any eigenphase $\varepsilon$ at $\mathbf{k}$, there exists an eigenphase $-\varepsilon$ at $-\mathbf{k}$.
Given the off-diagonal matrix structure of the Ho-Chalker operator, there is an extra symmetry present, the twofold phase rotation symmetry, which rotates the eigenphase $\varepsilon$ to $\varepsilon+\pi$ via
\begin{eqnarray}
\mathcal{U}_{pr}{\cal S}(\mathbf{k})\mathcal{U}_{pr}^\dagger = -{\cal S}(\mathbf{k}),
\end{eqnarray}
with $\mathcal{U}_{pr}=\tau_z\otimes \mathbb{I}_{4\times 4}$.
As a result, the eigenphases of the network are fully encoded in the fundamental domain $\varepsilon \in [-\pi/2,\pi/2)$, which reduces the eight-band network model to an effectively four-band model. 
Further, due to the $C_4\mathcal{T}$ symmetry, the system is doubly degenerate at high symmetry points $\mathbf{k}=(0,0)$ and $(\pi,\pi)$.

\section{higher-order topological phase}
\label{sec:HOTP}
 
Observing corner modes requires a finite-sized system, for which we need to specify a termination. 
For network models, it corresponds to a hard-wall boundary where incident modes are fully reflected, i.e., at the top (bottom) edge the scattering can only happen between modes $\psi_{1,\sigma}$ ($\psi_{2,\sigma}$) and $\psi_{4,\sigma}$ ($\psi_{3,\sigma}$).
With respect to the global $C_4\mathcal{T}$ symmetry, the edge scattering nodes take the form, 
\begin{eqnarray}
S_{1\text{e}}&=&-\cos\alpha_{\text{e}}\tau_0\otimes\tau_0-i\sin\alpha_{\text{e}}\tau_y\otimes\tau_x,\\
S_{2\text{e}}&=&-\cos\alpha_{\text{e}}\tau_0\otimes\tau_0+i\sin\alpha_{\text{e}}\tau_0\otimes\tau_y,
\end{eqnarray}
with $\alpha_{\text{e}}$ being the edge scattering angle (see Fig.~\ref{fig:model}(b)). 

To give an intuitive picture of the in-gap corner states, we refer to the decoupled limit with all scattering nodes being either fully reflecting or fully transmitting~\cite{Liu2020-hoti}.
The HOTP limit with $(\theta_1,\theta_2,\alpha,\beta)=(0.5\pi,0,0,0)$ and $\alpha_{\text{e}}=0$ is then shown in Fig. \ref{fig:model}(b), in which all bulk modes and edge modes are trapped in $4$-step-evolution $\pi$-flux circles (black and red arrows), which give eigenphases at $\pm\pi/4$ and $\pm 3\pi/4$.
For corner loops, propagating modes carrying no phases return back after a $2$-step evolution and result in a total of 8 corner modes at $0$ eigenphase, respectively [see Fig.~\ref{fig:model}(c)]. 

The in-gap corner states are protected by a global $C_4\mathcal{T}$ symmetry, meaning that the resulting phase is an intrinsic HOTP~\cite{Trifunovic_intrinsic}, which is independent of the choice of boundary conditions.
To reveal this, we focus on the top boundary of the HOTP limit, which is already decoupled from the bulk and consists of two types of reduced $2\times 2$ scattering nodes, 
\begin{eqnarray}
S_{1,\text{top}}=-\cos\alpha_{\text{e}}\tau_0-i\sin\alpha_{\text{e}}\tau_y,~S_{2,\text{top}}=i\tau_y.\label{eq:boundary_setting}
\end{eqnarray} 
This is reminiscent of the one-dimensional Su-Schrieffer-Heeger chain with $S_{1,\text{top}}$ and $S_{2,\text{top}}$ corresponding to the ``intracell'' and the ``intercell'' coupling, respectively~\cite{ssh}. 
In this language, $S_{2,\text{top}}$ representing a full transmission can be regarded as the strongest intercell coupling, which forces the top boundary system to be topologically nontrivial, except a gap closing when $\alpha_{\text{e}}=0.5\pi$ (where $S_{1,\text{top}}=-S_{2,\text{top}}$). 
Consequently, as long as $C_4\mathcal{T}$ symmetry is preserved these intrinsic corner modes cannot be removed by varying the edge scattering angle $\alpha_{\text{e}}$, even with an edge gap closing and reopening.  

\section{phase diagram}
\label{sec:phase_diagram}

\begin{figure}
	\centering
	\includegraphics[width=\linewidth]{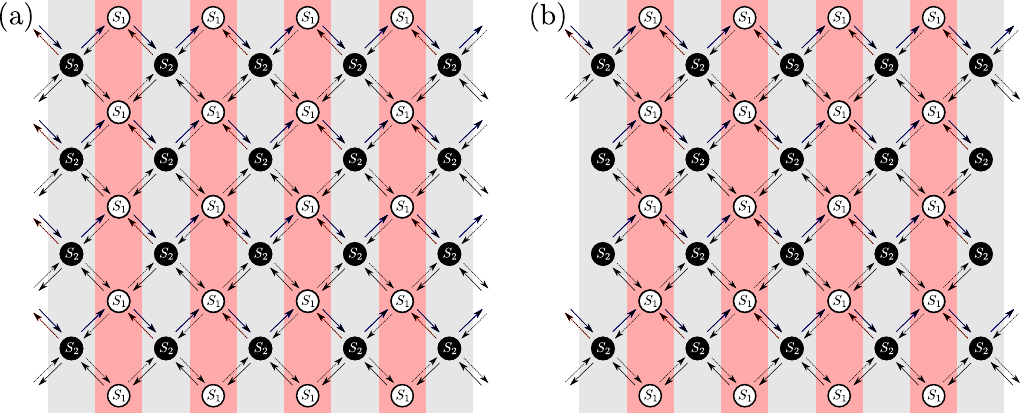}
	\caption{The left and right panels are the two-terminal and four-terminal transport setting, respectively. 
Here, the red and gray region indicate the scattering slices composed of by $S_1$ and $S_2$, respectively.
\label{fig:transport_setting}}
\end{figure}

Since the $C_4\mathcal{T}$ symmetry operation enforces pairs of counter-propagating modes at the same edge, the system cannot support a strong topological phase, i.e., the Chern insulator phase with chiral edge states. 
In this sense, except the HOTP, another possible nontrivial state of a two-dimensional $C_4\mathcal{T}$ system is the WTP.

Our first tool to distinguish these phases is transport simulations. 
An advantage of network models is that, without generating a full Hamiltonian matrix (or Floquet operator), their scattering matrices can be obtained by the so-called Redheffer star product~\footnote{P. W. Brouwer, Ph.D. thesis, Leiden University, 1997.}, which allows to evaluate the transport properties slice by slice,

\begin{align}\label{eq:combiningS}
\begin{split}
& r_C = r_L + t_L' r_R (1 - r_L' r_R)^{-1} t_L, \\
& t_C = t_R (1 - r_L' r_R)^{-1} t_L, \\
& t_C' = t_L' (1 - r_R r_L')^{-1} t_R', \\
& r_C' = r_R' + t_R r_L' (1 - r_R r_L')^{-1} t_R',
\end{split}
\end{align}
with $S_{R,L}=\begin{pmatrix}
r_{R,L}&t'_{R,L}\\t_{R,L}&r_{R,L}
\end{pmatrix}$ being the scattering matrix for the left and right verticle slice, respectively. 
Here, each slice is composed of either $S_1$ or $S_2$ scattering nodes (see Fig.~\ref{fig:transport_setting}).

\begin{figure}
\centering
\includegraphics[width=\linewidth]{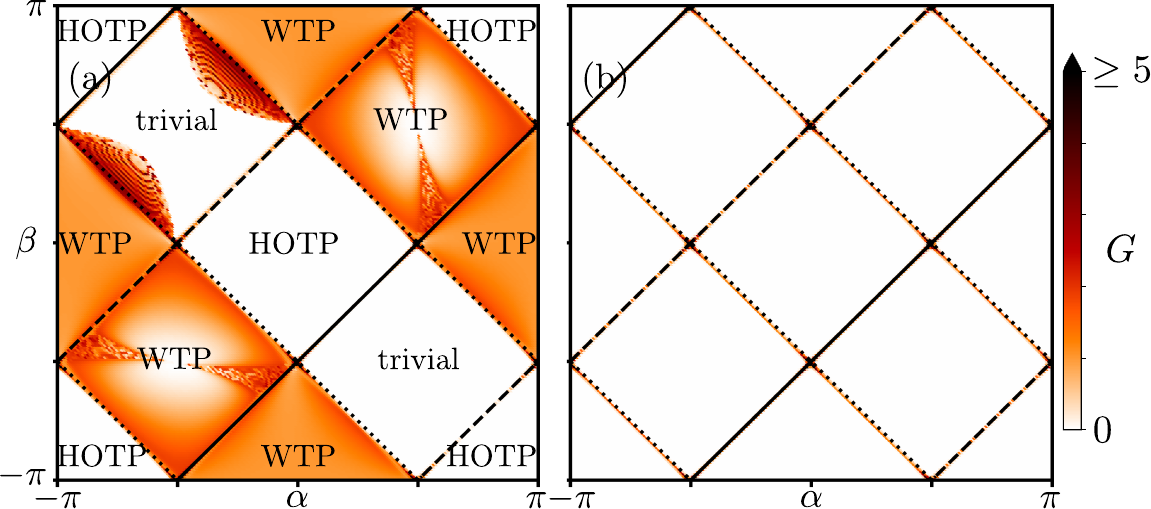}
\caption{Panels (a) and (b) are the two-terminal transmission as a function of $\alpha$ and $\beta$ with OBCs and PBCs at phase $\varepsilon=0$, respectively.  
Here, the solid, the dashed, and the dotted lines correspond to the phase boundaries from Eq.~\eqref{eq:phase_boundary}, respectively.
All plots are with $(\theta_1,~\theta_2)=(0.5\pi,~0)$, $\alpha_{\text{e}}=0.25\pi$. The transmission is calculated with a system size $50\times 50$ unit cells.
\label{fig:two_terminal_transmission}
}
\end{figure}

\begin{figure}
\centering
\includegraphics[width=\linewidth]{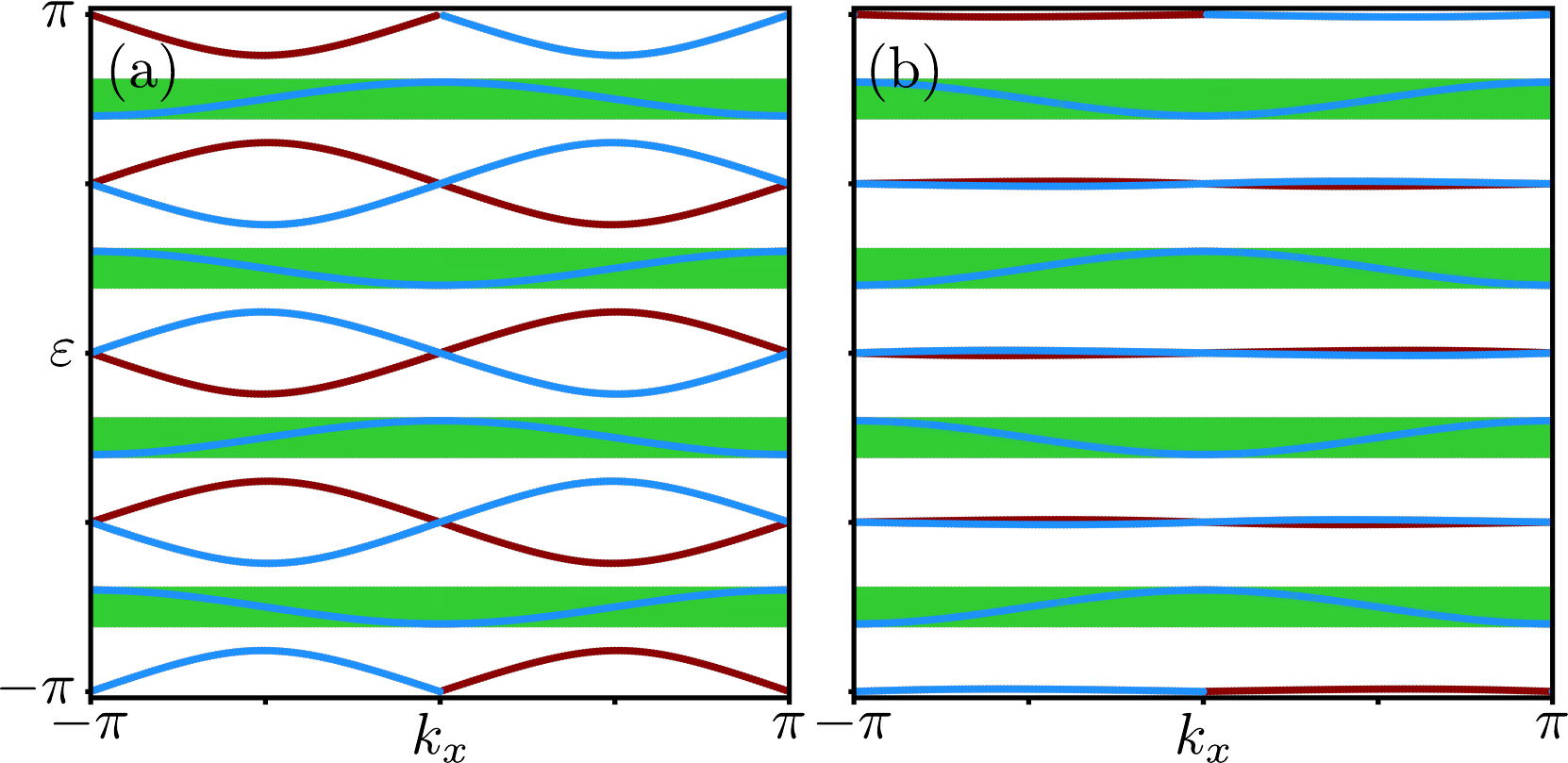}
\caption{The ribbon geometry spectrum of the two different types of weak topological phases with $(\alpha, \beta)=(0, 0.9\pi)$ (a) and $(0.5\pi, 0.4\pi)$ (b), respectively. 
Here, we consider a system infinite along the horizontal direction and consisting of 20 unit cells in the vertical direction. The colors represent the position of individual states, with red/blue indicating modes localized on the top/bottom boundaries and green indicating bulk modes. All plots are with $(\theta_1,\theta_2)=(0.5\pi, 0)$.
\label{fig:2wtp}
}
\end{figure}

The two-terminal transmission is shown in Fig.~\ref{fig:two_terminal_transmission}(a-b). Here we attach leads at the left and right boundary, and apply various boundary conditions along the $y$-direction. 
We only focus on the transmission at eigenphase $\varepsilon = 0$, since the system is protected by particle-hole symmetry. 
As the strong topological phase is forbidden, a nonzero transmission with open boundary conditions (OBC) only implies either a WTP or a gap closure.  
Here, the WTP indicates in-gap counter-propagating edge states without connecting to the bulk bands [see Fig.~\ref{fig:2wtp}(a) and (b)].
Further applying periodic boundary conditions (PBC) to exclude the contribution from the counter-propagating edge states of the WTP, we fully identify all phases. 

The phase transitions that we obtain numerically have a one-to-one correspondence to the bulk gap closing, which can be determined analytically by the eigenvalue equation $\text{Det}[\mathcal{S}(\mathbf{k})-\mathbb{I}_{8\times 8}]=0$. 
It provides
\begin{eqnarray}
\alpha-\beta&=&2n\pi-\frac{\pi}{2},~\text{for}~\mathbf{k}=(0,0),\nonumber\\
\alpha-\beta&=&2n\pi+\frac{\pi}{2},~\text{for}~\mathbf{k}=(\pi,\pi),\\
\alpha+\beta&=&2n\pi+\pi,~\text{for}~\mathbf{k}=(0,\pi)~\text{and}~(\pi,0),\nonumber
\label{eq:phase_boundary}
\end{eqnarray}
with $n\in\mathbb{Z}$ when $(\theta_1,\theta_2)=(0.5\pi, 0)$. 

We found the phase transition between the HOTP and the trivial phase occurs only at either $\mathbf{k}=(0,0)$ or $(\pi,\pi)$, and the phase transition between the WTP and the HOTP (or the trivial) happens at $\mathbf{k}=(0,\pi)$ and $(\pi,0)$ related by $C_4\mathcal{T}$ symmetry.
Furthermore, there is a bulk gap closing between WTPs (at high symmetry points), which we later identify as a topological phase transition in Sec.~\ref{sec:top_inv}. 
Note that, due to the vanishing group velocity of counter-propagating modes at $(\alpha_1,\beta_1)=\pm(\pi/2,\pi/2)$, the corresponding transmission becomes zero [see Fig.~\ref{fig:two_terminal_transmission}(a) and Fig.~\ref{fig:2wtp}(b)].

\section{topological invariants}
\label{sec:top_inv}

Our second tool to identify topological phases of matter is topological invariants.

\subsection{Reflection matrix invariant}

\begin{figure}
\centering
\includegraphics[width=\linewidth]{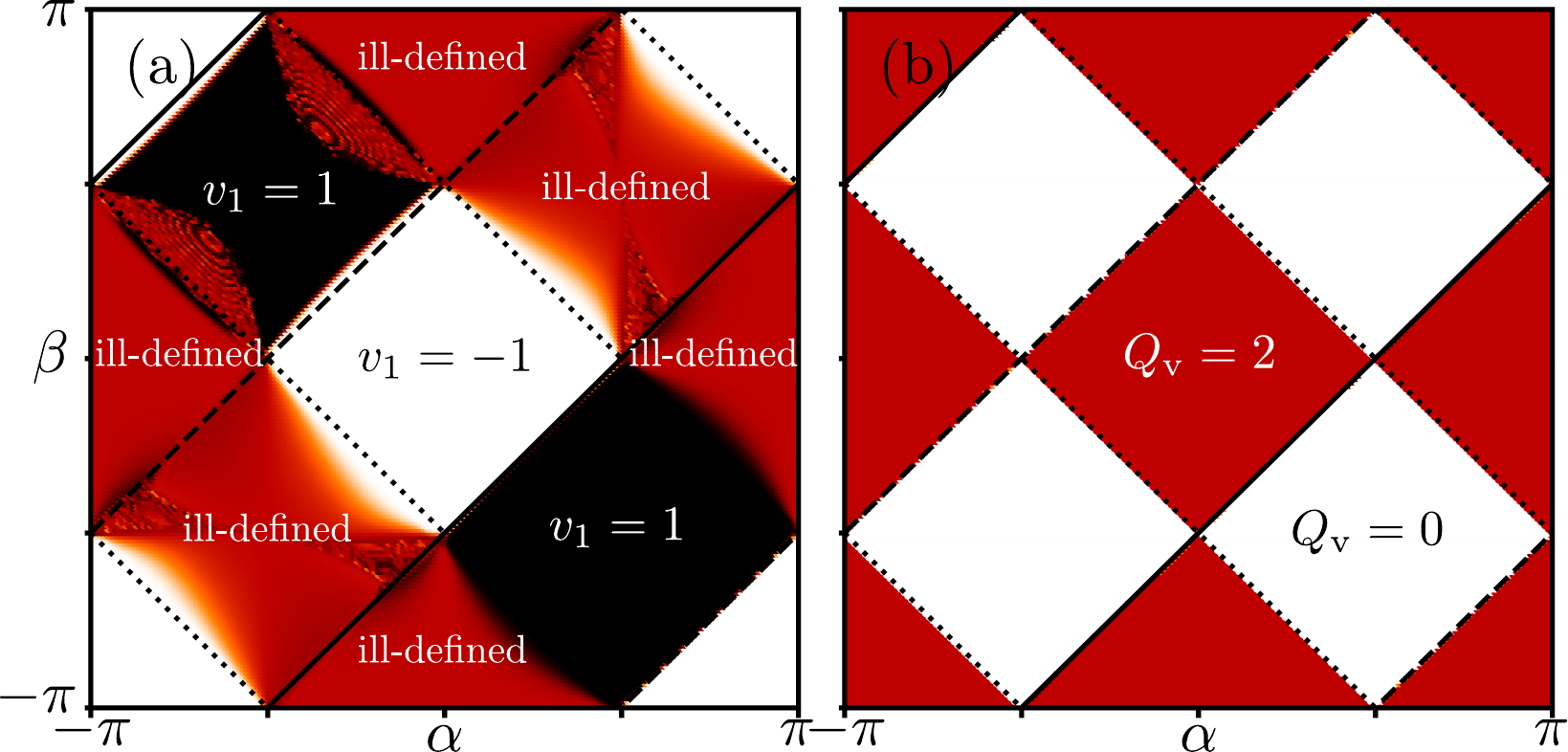}
\caption{Panels (a) and (b) are the reflection matrix invariant and the Pffafian invariant as a function of $\alpha$ and $\beta$, respectively. The solid, the dashed, and the dotted lines are the analytical phase boundaries. All plots are with $(\theta_1,~\theta_2)=(0.5\pi,~0)$. Here, for the reflection matrix invariant we set $\alpha_{\text{e}}=0.25\pi$, and use the same system size as in Fig.~\ref{fig:two_terminal_transmission}.
\label{fig:topological_invariant}
}
\end{figure}

Since we are dealing with scattering arrays, the reflection matrix invariant from the four-terminal geometry scattering setting serves as a direct topological signature.
Here, the four leads are only attached at the four corner unit cells, see Fig.~\ref{fig:transport_setting}(b).
The scattering matrix is given by
\begin{eqnarray}
S_{4-\text{terminal}}=\begin{pmatrix}r_{11}&t_{12}&t_{13}&t_{14}\\t_{21}&r_{22}&t_{23}&t_{24}\\t_{31}&t_{32}&r_{33}&t_{34}\\t_{41}&t_{42}&t_{43}&r_{44}
\end{pmatrix},
\end{eqnarray}
with $r_{ii}$ being the reflection matrix of the $i$-th terminal and $t_{ij}$ being the transmission matrix from the $j$-th terminal to $i$-th terminal.
For a gapped system, the topological invariant is defined as the sign of the determinant of the $2\times 2$ unitary reflection matrix $r_{ii}$ ($i=1$, $2$, $3$, $4$), which is guaranteed to be real by particle-hole symmetry. 
Thus, when both the bulk and edges are gapped around $\varepsilon=0$, the reflection matrix determinant establishes a $\mathbb{Z}_2$ invariant to distinguish the HOTP and the trivial phase by~\cite{scattering_invariant}
\begin{eqnarray}
v_{i}=\text{sgn}(\text{det}[r_{ii}]).
\end{eqnarray}
The HOTP and the trivial phase correspond to $v_{i}=-1$ and $1$, respectively [see Fig. \ref{fig:topological_invariant}(a)].
In the WTP, the edge gap is closed, which results in a zero reflection matrix for which the $\mathbb{Z}_2$ invariant is no longer well-defined.
Note that, the mismatch in the phase boundary between the WTP and the trivial phase is an artifact of the choice of $\alpha_\text{e}$, which can be eliminated by setting $\alpha_{\text{e}}\rightarrow 0$, where couterpropagating modes are fully gapped out in the trivial phase.

\subsection{Pfaffian invariant}

\begin{figure}
	\centering
	\includegraphics[width=0.5\linewidth]{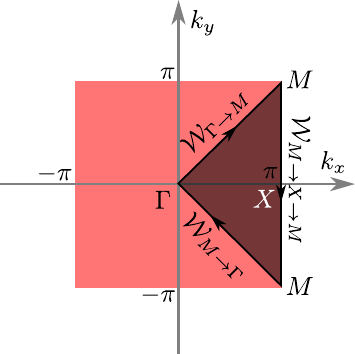}
	\caption{Sketch of the first Brillouin zone and the effective Brillouin zone (dark region). The black solid line is the path of the Wilson loop. Here, $\mathcal{W}_{\Gamma\rightarrow M}$ is equivalent to $\mathcal{W}_{M\rightarrow \Gamma}$ by $\mathcal{C}_4\mathcal{T}$ symmetry, and $\mathcal{W}_{M\rightarrow X}$ is related to $\mathcal{W}_{X\rightarrow M}$ by $\mathcal{C}_2=(\mathcal{C}_4\mathcal{T})^2$.
\label{fig:effective_BZ}}
\end{figure}

Another invariant is the Pfaffian invariant defined for the bulk~\cite{pfaffian}.
For systems with spatial symmetries, such as rotation or mirror, the nontrivial band inversions can be identified by symmetry indicators at high symmetry momenta~\cite{Po2020, Po2017a}.
In systems with spinful time reversal symmetry, like the quantum spin Hall effect~\cite{qsh2, qsh1}, due to Kramers' degeneracy, when computing the topological invariant we need to take the global gauge of wavefunction into account~\cite{tr1, tr2}.
To bypass this, an efficient approach is to use the Pfaffian of the sewing matrix~\cite{tr1, tr2}, 
\begin{eqnarray}
w_{nm}(\mathbf{k})=\langle \psi_{n}(\mathbf{k})|\mathcal{U}_{\mathcal{T}}\mathcal{K}|\psi_m(-\mathbf{k})\rangle.
\end{eqnarray} 
which is skew-symmetric to respect time reversal symmetry. 
However, such a shortcut is absent with $C_4\mathcal{T}$ symmetry, as in general $\langle \psi_{n}(\mathbf{k})|\mathcal{U}_{C_4\mathcal{T}}\mathcal{K}|\psi_n(-C_4\mathbf{k})\rangle\neq0$. 
But alternatively, it is still possible to define a generalized Pfaffian $\text{pf}(w(\mathbf{k}))$ to get rid of the gauge-dependence, as detailed in Ref.~\cite{Daniel_pfaffian}. 
Afterwards, analogous to quantum spin-Hall systems, the topological invariant for an occupied band for a  $C_4\mathcal{T}$ symmetric system reads~\cite{pfaffian}
\begin{eqnarray}
\hspace{-5mm}Q_{\text{occ}}=\frac{1}{\pi}\text{tr} \left( \int_{\text{EBZ}}d\mathbf{k}^2\mathcal{F}_{\text{occ}}-2i\text{log}\tilde{\mathcal{W}}^{\text{occ}}_{\Gamma\rightarrow M} \right) \text{mod}~4.
\label{eq:z4}
\end{eqnarray}
Here, the effective Brillouin zone (EBZ) is formed by a closed path through high symmetry momenta $\Gamma=(0, 0)$, $X=(0,0)$, and $M=(\pi,\pi)$ in the first Brillouin zone, see Fig.~\ref{fig:effective_BZ}. $\mathcal{F}_{\text{occ}}$ is the Berry curvature of the occupied band at $\mathbf{k}$, and $\tilde{\mathcal{W}}^{\text{occ}}_{\Gamma\rightarrow M}=\text{pf}(w(M))^{-1}\text{det}(\mathcal{W}_{\Gamma\rightarrow M})\text{pf}(w(\Gamma))$ is the corresponding dressed Wilson line with $\mathcal{W}_{\Gamma\rightarrow M}=\Psi(M)\prod_{\mathbf{k}\in\Gamma\rightarrow M}\Psi^{*}(\mathbf{k})\Psi(\mathbf{k})\Psi^{*}(\Gamma)$ being the conventional Wilson line ($\Psi=(\psi_1,\psi_2,\cdots,\psi_n)$ is the eigenstate set for the occupied band). 
In the whole calculation, $\tilde{\mathcal{W}}_{M\rightarrow X\rightarrow M}=\mathcal{W}_{M\rightarrow X\rightarrow M}$ is excluded as it is quantized by the twofold rotation symmetry $\mathcal{C}_2=(\mathcal{C}_4\mathcal{T})^2$~\cite{pfaffian}.

In our network model, each eigenphase band is doubly degenerate at $\Gamma$ and $M$ due to the $\mathcal{C}_4\mathcal{T}$ symmetry. 
In one fundamental domain, it reduces the network to a two-band model. 
In this sense, there are only two independent $Q$ indices, which we label as $Q_\text{v}$ for the band at $\varepsilon<0$ and $Q_\text{c}$ for the band at $\varepsilon>0$, in analogy with valence and conduction bands.
The presence of particle-hole symmetry further leads to $Q_\text{v}=-Q_\text{c}$ (see App.~\ref{sec:phs_z4} for more details).
We then focus on $Q_{\text{v}}$. 
The phase diagram shown in Fig.~\ref{fig:topological_invariant}(b) suggests there are two distinct topological phases with $Q_\text{v}=2$ and $Q_\text{v}=0$. 
Together with the reflection matrix invariant as well as two-terminal transmission, we fully distinguish all phases: $Q_\text{v}=2$ for the HOTP and one type of WTP, and $Q_\text{v}=0$ for the trivial phase and the second type of WTP. 
Transitions between different topological phases correspond to a bulk gap closing, which has been verified by the eigenvalue equation shown in Sec.~\ref{sec:phase_diagram}. 

We note that, in general, $Q$ can take on the values $0$, $1$, $2$, or $3$. 
For the parameter setting $(\theta_1, \theta_2)=(0.5\pi,0)$, the phases with $Q=1$ and $Q=3$ are absent. 
However, this absence is not a limitation of our network model or the result of a symmetry restriction. As demonstrated in the App. \ref{sec:pd_Q13}, these phases do appear in other parameter regimes.

\section{Hidden $\pi/2$ corner states}
\label{sec:hidden}

\begin{figure*}
	\centering
	\includegraphics[width=\linewidth]{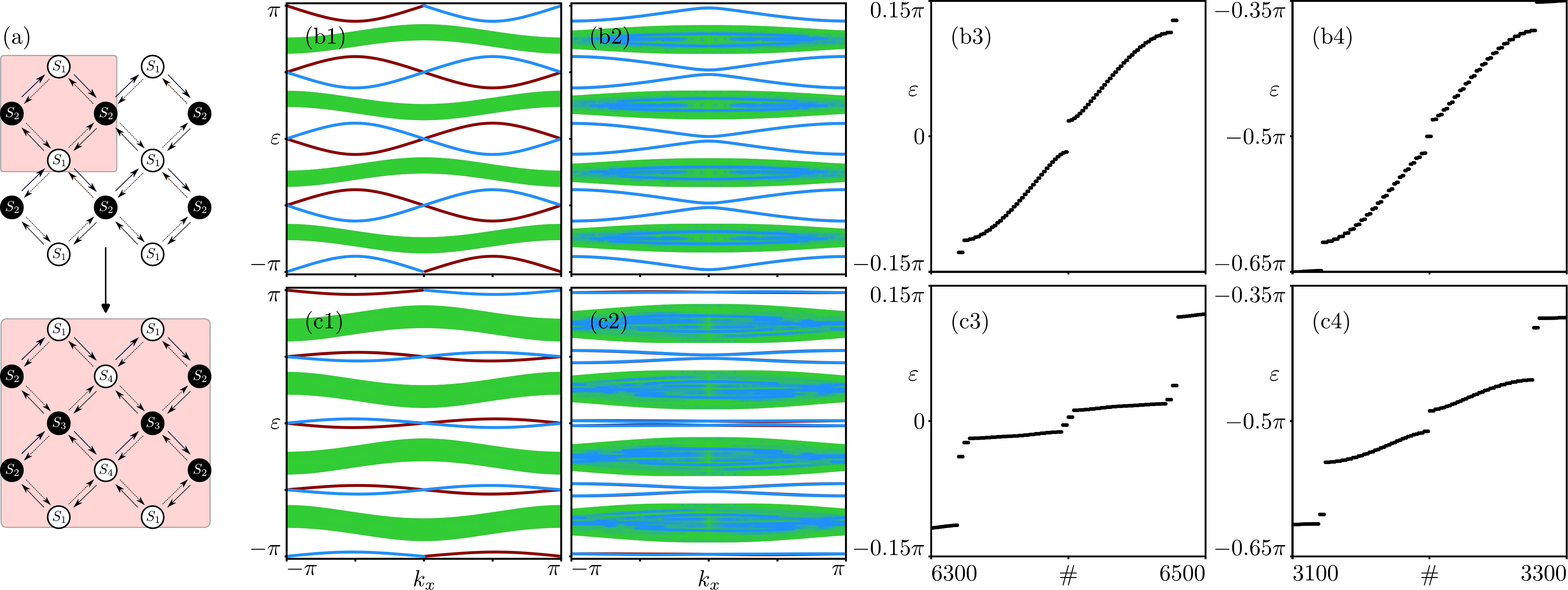}
	\caption{ (a) The left panel is the original $C_4\mathcal{T}$-symmetric network with $2\times 2$ unit cells (shaded area) and the right panel is the unit cell of their corresponding doubled network with symmetry preserving perturbation.
(b1) and (c1) are the ribbon geometry spectrum of the original network for the $Q_\text{v}=2$ with $(\alpha, \beta)=(0.1\pi, 0.9\pi)$ and the $Q_\text{v}=0$ WTP with $(\alpha, \beta)=(0.4\pi, 0.65\pi)$, respectively. 
(b2) and (c2) are the ribbon geometry spectrum and (b3-c3) and (b4-c4) are the zoom-in finite-sized spectrum near $0$ and $\pi/2$ eigenphase for their corresponding doubled network, respectively. All plots are with $(\theta_1, \theta_2)=(0.5\pi, 0)$, $\delta=0.1\pi$, and $\alpha_{\text{e}}=0.25\pi$.
Here, we use a system size $60$ unit cells and a system size $40\times 40$ unit cells for the ribbon geometry and finite-sized eigenphase spectrum, respectively.
\label{fig:hidden_hoti}
}
\end{figure*}

$Q_\text{v}=2$ suggests the first type of WTP is a HOTP, however, the corresponding corner modes vanish due to the conducting edges (due to the counter-propagating boundary modes). 
To reveal these corner states, we double the unit cell of the original network and introduce a $C_4\mathcal{T}$ symmetry preserving perturbation, as sketched in Fig.~\ref{fig:hidden_hoti}(a), to gap out these translational-symmetry protected boundary modes. 
In the modified network, the rotation center remains in the middle of the unit cell, relating $S_1$ ($S_3$) nodes to $S_2$ ($S_4$) nodes. 
The perturbation $\delta$ is applied to $S_3$ and $S_4$ with
\begin{eqnarray}
S_3=S_2(\theta_1, \theta_2+\delta, \alpha,\beta),~S_4=S_1(\theta_1, \theta_2+\delta, \alpha,\beta).~ 
\end{eqnarray}
The $C_4\mathcal{T}$ symmetry is maintained as $S_3$ and $S_4$ inherit the symmetry constraint in Eq.~\eqref{eq:sym_node} as well.
When $\delta=0$ the system returns back to the original network. 
A nonzero $\delta$ will dimerize the doubled network due to the difference between the intracell and the intercell coupling. 
It breaks the translational symmetry and gaps out the counter-propagating edge states. 
In this procedure, a small $\delta$ can only open the edge gap and still preserves the bulk gap, shown in Fig.~\ref{fig:hidden_hoti}(b1-b2) and (c1-c2).
Afterwards, the hidden corner states are now present in the finite-sized system for the $Q_\text{v}=2$ WTP  (see Fig.~\ref{fig:hidden_hoti}(b3-b4)). 
As a comparison, for the $Q_\text{v}=0$ WTP, gapping out the counter-propagating edge states does not lead to corner states [see Fig.~\ref{fig:hidden_hoti}(c3-c4)].

Quite interestingly, the corner modes now have $\pm \pi/2$ eigenphase, instead of $0$ or $\pi$ eigenphase. 
This is a consequence of the combined effect of particle-hole symmetry and twofold phase-rotation symmetry. 
In topological superconducting systems, the fundamental particle-hole symmetry pins the Majorana corner modes to $0$ energy ($\varepsilon=-\varepsilon$). 
The presence of twofold phase-rotation symmetry leads to an extra constraint for the evolution operator,
\begin{eqnarray}
\mathcal{U}_{ph}\mathcal{U}_{pr}{\cal S}(\mathbf{k})(\mathcal{U}_{ph}\mathcal{U}_{pr})^\dagger=-
{\cal S}(-\mathbf{k}).
\end{eqnarray}
It maps an eigenstate $\Psi$ at momentum $\mathbf{k}$ and eigenphase $\varepsilon$ to the eigenstate $\mathcal{U}_{ph}\mathcal{U}_{pr}\Psi$ at momentum $-\mathbf{k}$ and eigenphase $-\varepsilon+\pi$,
This gives an additional option for symmetry protected topological phases,
\begin{eqnarray}
\varepsilon=-\varepsilon\pm\pi.
\end{eqnarray}
which provides Majorana corner modes with $\varepsilon=\pm\pi/2$. 
Without such a symmetry, these states can in principle shift away from $\pm\pi/2$ eigenphase in a pairwise fashion without breaking particle-hole symmetry. 

We emphasize that, these states are different from the recently introduced $\pm\pi/2$ eigenphase modes in an acoustic Floquet system, which are generated only at a fine tuning point~\cite{pi_over_2_modes}. 
Here, the presence of such states is originates from the phase-rotation symmetry, a symmetry only occuring in network models. 
Unlike Floquet systems, in general, only possessing two fundamental domains (i.e., in the presence of particle-hole symmetry, centers around $\varepsilon=0$ and $\pi$, respectively), a network model with $n$-fold phase-rotation symmetry can support $n$ such domains. 
For a topological superconductor, it means 
\begin{eqnarray}
\varepsilon = -\varepsilon+2m\pi/n, m\in\mathbb{Z}~\text{and}~|m|\leq n.
\end{eqnarray}
This provides a way to generate symmetry protected corner modes at arbitrary fractional eigenphases. 

It is worth mentioning that the Pfaffian index, originally designed for Hamiltonian systems, fails to distinguish these two distinct types of HOTPs, despite the existence of a phase transition between them. 
One possible explanation is that the Pfaffian index is defined to categorize inequivalent phases through distinct obstructed atomic insulating phases~\cite{pfaffian}. 
However, network models lack a proper definition of both Wannier centers and Wyckoff positions, limiting the capability to capture a more complex real-space configuration beyond the static scenarios. 
Another possible reason is that, a network model, featuring multiple bulk gaps to hold topological modes due to the phase-rotation symmetry, can facilitate a new type of topological phase, dubbed as ``anomalous topology''~\cite{Rudner2013}. 
In this case, the system possesses a trivial topological index, but still maintains topological boundary modes. 
A well-known  example is the anomalous Floquet topological insulator, where the bulk band exhibits zero Chern number, yet each gap harbors an equal number of chiral edge states. 
In this context, the emergence of $\pi/2$ modes might suggest the presence of such anomalous topology, characterized by a vanishing Pfaffian index. 
We plan to explore this direction in future research.

\section{Conclusion}
\label{sec:conc}
We have constructed a two-dimensional network model for magnetic higher-order topological phases. 
It provides a total of $8$ corner modes protected by a combination of phase-rotation symmetry, particle-hole symmetry, and $C_4\mathcal{T}$ symmetry. 
By evaluating both the reflection matrix invariant, the Pfaffian invariant, as well as the transport properties, we identified four distinct phases: $(Q_{\text{v}}, v_1)=(2, -1)$ for the HOTP, $(2, \text{ill-defined})$ for the hidden-HOTP WTP, $(0, \text{ill-defined})$ for the conventional WTP, and $(0, 1)$ for the trivial phase.
For the hidden-HOTP WTP, a $C_4\mathcal{T}$ symmetry-preserving dimerization opens the edge gap and reveals the corner modes at $\pm\pi/2$ eigenphase. 

For future research, there are several directions to explore. First, as a prototypical model possessing a combined space-time symmetry, our work opens a way towards magnetic topological phases and their associated localization-delocalization transitions using scattering arrays~\cite{Ozawa2019, Hafezi2011, Hafezi2013, Liang2013}. 
Secondly, the emergence of $\pi/2$ corner modes provides a valuable platform to study symmetry protected $2m\pi/n$ modes, which can be potentially used to realize parafermions, as suggested in Ref.~\cite{pi_over_2_modes}.
On a more practical level, scattering arrays as an experimentally accessible system have already been realized in both optical fibers and coupled ring resonators~\cite{exp2, exp3, Hu2015, Hafezi2013a, Gao2018, Gao2016, Hafezi2011a, ElHassan2019}. 
We hope our work will motivate new experiments to generate magnetic higher-order topological phases in future studies.

\begin{acknowledgments}

We thank Ulrike Nitzsche for technical assistance and Isidora Araya Day for useful discussions.
D.~V. was supported by the Swedish Research Council (VR), the Knut and Alice Wallenberg Foundation, and the National Research, Development and Innovation Office of Hungary under OTKA grant no. FK 146499.
This work was supported by the Deutsche Forschungsgemeinschaft (DFG, German Research Foundation) under Germany’s Excellence Strategy through the W\"{u}rzburg-Dresden Cluster of Excellence on Complexity and Topology in Quantum Matter – \textit{ct.qmat} (EXC 2147, project-id 390858490 and 392019).
A.G.M. thanks IFW Dresden for their warm hospitality and support during his stay, where this work was initiated.

\end{acknowledgments}

\appendix

\section{Phase diagram with $Q=1$ and $Q=3$}
\label{sec:pd_Q13}

\begin{figure}
\centering
\includegraphics[width=\linewidth]{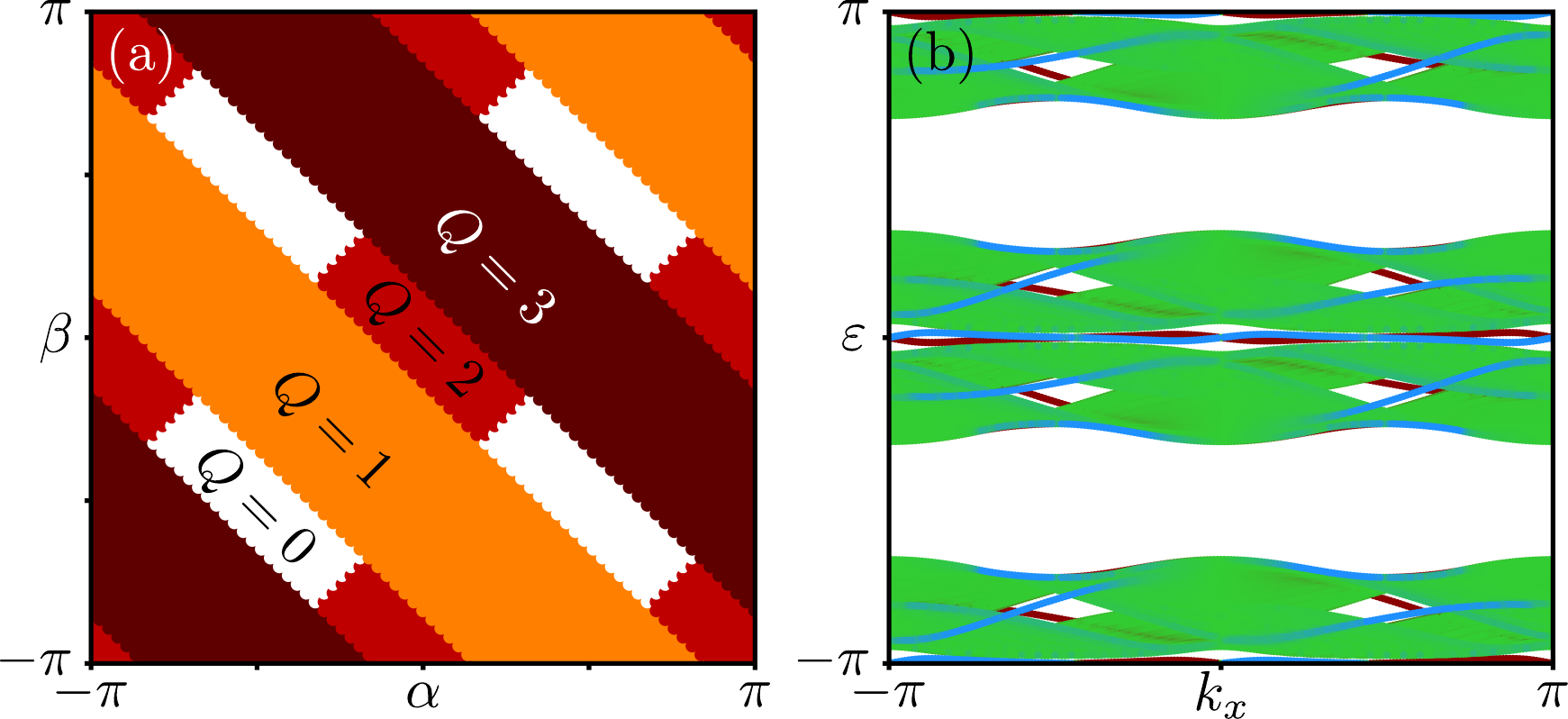}
\caption{Panels (a) and (b) are the phase diagram of Pfaffian invariant and the ribbon geometry spectrum of $Q=1$, respectively. All plots are with $(\theta_1,~\theta_2)=(0.4\pi,~0.5\pi)$. Here, for the spectrum plot, we use $(\alpha_1,\alpha_2)=(0.25\pi, 0.25\pi)$ and a system size $20$ unit cells.
\label{fig:topological_invariant1}
}
\end{figure}

In the parameter region with $(\theta_1, \theta_2)=(0.4\pi, 0.5\pi)$, we indeed observe the topological phases with the Pfaffian invariant $Q=1$ and $Q=3$ (see Fig. \ref{fig:topological_invariant1}(a)). 
We verify that both $Q=1$ and $Q=3$ correspond to the WTP. For simplicity, we show the ribbon geometry spectrum of the $Q=1$ case, in which the counter-propagating edge mode presents in the $\varepsilon=0$ gap, as shown in Fig. \ref{fig:topological_invariant1}(b).

\section{Interplay of $\mathbb{Z}_4$ index and particle-hole symmetry}
\label{sec:phs_z4}

The definition of $\mathbb{Z}_4$ index only depending on $C_4\mathcal{T}$ symmetry implies that each energy band will be assigned an independent topological number ($Q=0,~1,~2,~3$), since it relates an eigenstate $\psi$ at $E$ to an eigenstate $\mathcal{U}_{C_4\mathcal{T}}\mathcal{K}\psi$ at the same energy,
\begin{eqnarray}
(k_x,k_y,E,\psi)\rightarrow (k_y,-k_x,E,\mathcal{U}_{C_4\mathcal{T}}\mathcal{K}\psi).~\label{eq:c4trs}
\end{eqnarray}
When adapting $\mathbb{Z}_4$ index to a topological superconducting system, the presence of particle-hole symmetry provides another restriction,
\begin{eqnarray}
(k_x,k_y,E,\psi)\rightarrow (-k_x,-k_y,-E,\mathcal{U}_{\mathcal{P}}\mathcal{K}\psi).~\label{eq:ph}
\end{eqnarray}
Here, $\mathcal{U}_{\mathcal{P}}$ is a unitary matrix of particle-hole symmetry $\mathcal{P}$. 
In this sense, for a $C_4\mathcal{T}$-symmetric superconducting system, the $\mathbb{Z}_4$ indexes for the particle-hole symmetric bands are not independent, specifically, 
\begin{eqnarray}
Q(E)=-Q(-E).
\end{eqnarray}

To reach this relation, we start from the mathematical definition of $\mathbb{Z}_4$ index in Eq.~\eqref{eq:z4}. 
It contains two parts, the first part is an integral of Berry curvature. 
In a discretized $N_x\times N_y$ Brillouin zone, the Berry curvature can be expressed as~\cite{Fukui_chern},
\begin{eqnarray}
\tilde{F}_{12}(\mathbf{k})&\equiv&\text{ln}~U_1(\mathbf{k})U_2(\mathbf{k}+\hat{1})U_1(\mathbf{k}+\hat{2})^{-1}U_2(\mathbf{k})^{-1},\nonumber\\
&&-\pi<\frac{1}{i}\tilde{F}_{12}(\mathbf{k})\leq \pi.
\end{eqnarray}
Here, $U_{\mu}(\mathbf{k})=\text{det}~\Psi^\dagger(\mathbf{k})\Psi(\mathbf{k}+\hat{\mu})/\mathcal{N}_\mu(\mathbf{k})$ with $\mathcal{N}_\mu(\mathbf{k})=|\text{det}~\Psi^\dagger(\mathbf{k})\Psi(\mathbf{k}+\hat{\mu})|$, $\hat{\mu}=\hat{1}$ (or $\hat{2}$) represents the $k_x$ (or $k_y$) direction in momentum space with a magnitude $2\pi/N_x$ (or $2\pi/N_y$), and $\Psi=(\psi_1,\psi_2,\cdots,\psi_n)$ is a multiplet represented by the eigenstates of the system. 
Based on Eq.~\eqref{eq:c4trs}, operating it twice will give
\begin{eqnarray}
(k_x,k_y,E,\psi)\rightarrow (-k_x,-k_y,E,\mathcal{U}_{C_4\mathcal{T}}^2\psi).
\end{eqnarray}
Then we have $U_{\mu}(\mathbf{k}, E)=U_{\mu}(-\mathbf{k}, E)$, which leads to 
\begin{eqnarray}
\tilde{F}_{12}(\mathbf{k},E)=\tilde{F}_{12}(-\mathbf{k},E).~\label{eq:f_c4t}
\end{eqnarray}
Next, due to the anti-unitarity of particle-hole symmetry, one can obtain,
\begin{eqnarray}
\Psi^\dagger(-\mathbf{k})\Psi(-\mathbf{k}-\hat{\mu})&=&\text{det}~\Psi^{T}(\mathbf{k})\mathcal{U}_{\mathcal{P}}^\dagger\mathcal{U}_{\mathcal{P}}\Psi^{*}(\mathbf{k}+\hat{\mu})\nonumber\\
&=&(\Psi^\dagger(\mathbf{k})\Psi(\mathbf{k}+\hat{\mu}))^*.
\end{eqnarray}
This further provides,
\begin{eqnarray}
\tilde{F}_{12}(\mathbf{k},E)=-\tilde{F}_{12}(-\mathbf{k},-E).~\label{eq:f_ph}
\end{eqnarray}
Combining Eq.~\eqref{eq:f_c4t} and Eq.~\eqref{eq:f_ph}, we have
\begin{eqnarray}
\tilde{F}_{12}(\mathbf{k},E)=-\tilde{F}_{12}(\mathbf{k},-E).
\end{eqnarray}

The second part of $\mathbb{Z}_4$ index is the dressed Wilson loop, which is composed of three components, the conventional Wilson loop from $\Gamma$ to $M$ and the corresponding generalized Pfaffians at $\Gamma$ and $M$. 
For the Wilson loop $\mathcal{W}_{\Gamma\rightarrow M}$, 
\begin{eqnarray}
\mathcal{W}_{\Gamma\rightarrow M}=\Psi(M)\prod_{\mathbf{k}\in\Gamma\rightarrow M}\Psi^{*}(\mathbf{k})\Psi(\mathbf{k})\Psi^{*}(\Gamma),
\end{eqnarray}
combining particle-hole symmetry and $C_4\mathcal{T}$ symmetry will give,
\begin{eqnarray}
\mathcal{W}_{\Gamma\rightarrow M}(E)=\mathcal{W}_{\Gamma\rightarrow M}^{*}(-E).
\end{eqnarray}
For the generalized Pfaffian at high symmetry point, the sew matrix for an occupied band is defined as 
\begin{eqnarray}
\mathcal{U}_{s}=\Psi^\dagger \mathcal{U}_{C_4\mathcal{T}}\mathcal{K}\Psi.
\end{eqnarray}
Then the particle-hole partner band would be
\begin{eqnarray}
\mathcal{U}_{s}^{'}=\Psi^T\mathcal{U}_{\mathcal{P}}^\dagger \mathcal{U}_{C_4\mathcal{T}}\mathcal{U}_{\mathcal{P}}^{*}\Psi.
\end{eqnarray}
In our network model, $\mathcal{U}_{\mathcal{P}}=1$ and $\mathcal{U}_{C_4\mathcal{T}}=\mathcal{U}_{C_4\mathcal{T}}^{*}$, which then gives,
\begin{eqnarray}
\mathcal{U}_{s}^{'}=\mathcal{U}_{s}^{*}.
\end{eqnarray}
This makes the Pffafian satisfy 
\begin{eqnarray}
\text{pf}(w(\mathbf{k}_{\Gamma,M}, E))=\text{pf}(w(\mathbf{k}_{\Gamma,M}, -E))^{*}. 
\end{eqnarray}
Substitute these to Eq.~\eqref{eq:z4}, we finally reach $Q_\text{v}=-Q_\text{c}$

\bibliography{Refs}

\begin{thebibliography}{78}%
\makeatletter
\providecommand \@ifxundefined [1]{%
 \@ifx{#1\undefined}
}%
\providecommand \@ifnum [1]{%
 \ifnum #1\expandafter \@firstoftwo
 \else \expandafter \@secondoftwo
 \fi
}%
\providecommand \@ifx [1]{%
 \ifx #1\expandafter \@firstoftwo
 \else \expandafter \@secondoftwo
 \fi
}%
\providecommand \natexlab [1]{#1}%
\providecommand \enquote  [1]{``#1''}%
\providecommand \bibnamefont  [1]{#1}%
\providecommand \bibfnamefont [1]{#1}%
\providecommand \citenamefont [1]{#1}%
\providecommand \href@noop [0]{\@secondoftwo}%
\providecommand \href [0]{\begingroup \@sanitize@url \@href}%
\providecommand \@href[1]{\@@startlink{#1}\@@href}%
\providecommand \@@href[1]{\endgroup#1\@@endlink}%
\providecommand \@sanitize@url [0]{\catcode `\\12\catcode `\$12\catcode
  `\&12\catcode `\#12\catcode `\^12\catcode `\_12\catcode `\%12\relax}%
\providecommand \@@startlink[1]{}%
\providecommand \@@endlink[0]{}%
\providecommand \url  [0]{\begingroup\@sanitize@url \@url }%
\providecommand \@url [1]{\endgroup\@href {#1}{\urlprefix }}%
\providecommand \urlprefix  [0]{URL }%
\providecommand \Eprint [0]{\href }%
\providecommand \doibase [0]{https://doi.org/}%
\providecommand \selectlanguage [0]{\@gobble}%
\providecommand \bibinfo  [0]{\@secondoftwo}%
\providecommand \bibfield  [0]{\@secondoftwo}%
\providecommand \translation [1]{[#1]}%
\providecommand \BibitemOpen [0]{}%
\providecommand \bibitemStop [0]{}%
\providecommand \bibitemNoStop [0]{.\EOS\space}%
\providecommand \EOS [0]{\spacefactor3000\relax}%
\providecommand \BibitemShut  [1]{\csname bibitem#1\endcsname}%
\let\auto@bib@innerbib\@empty
%</preamble>
\bibitem [{\citenamefont {Hasan}\ and\ \citenamefont
  {Kane}(2010)}]{Kane2010RMP}%
  \BibitemOpen
  \bibfield  {author} {\bibinfo {author} {\bibfnamefont {M.~Z.}\ \bibnamefont
  {Hasan}}\ and\ \bibinfo {author} {\bibfnamefont {C.~L.}\ \bibnamefont
  {Kane}},\ }\bibfield  {title} {\bibinfo {title} {Colloquium: Topological
  insulators},\ }\href {https://doi.org/10.1103/RevModPhys.82.3045} {\bibfield
  {journal} {\bibinfo  {journal} {Rev. Mod. Phys.}\ }\textbf {\bibinfo {volume}
  {82}},\ \bibinfo {pages} {3045} (\bibinfo {year} {2010})}\BibitemShut
  {NoStop}%
\bibitem [{\citenamefont {Qi}\ and\ \citenamefont
  {Zhang}(2011)}]{Zhang2011RMP}%
  \BibitemOpen
  \bibfield  {author} {\bibinfo {author} {\bibfnamefont {X.-L.}\ \bibnamefont
  {Qi}}\ and\ \bibinfo {author} {\bibfnamefont {S.-C.}\ \bibnamefont {Zhang}},\
  }\bibfield  {title} {\bibinfo {title} {Topological insulators and
  superconductors},\ }\href {https://doi.org/10.1103/RevModPhys.83.1057}
  {\bibfield  {journal} {\bibinfo  {journal} {Rev. Mod. Phys.}\ }\textbf
  {\bibinfo {volume} {83}},\ \bibinfo {pages} {1057} (\bibinfo {year}
  {2011})}\BibitemShut {NoStop}%
\bibitem [{\citenamefont {Altland}\ and\ \citenamefont
  {Zirnbauer}(1997)}]{Altland1997}%
  \BibitemOpen
  \bibfield  {author} {\bibinfo {author} {\bibfnamefont {A.}~\bibnamefont
  {Altland}}\ and\ \bibinfo {author} {\bibfnamefont {M.~R.}\ \bibnamefont
  {Zirnbauer}},\ }\bibfield  {title} {\bibinfo {title} {Nonstandard symmetry
  classes in mesoscopic normal-superconducting hybrid structures},\ }\href
  {https://doi.org/10.1103/PhysRevB.55.1142} {\bibfield  {journal} {\bibinfo
  {journal} {Phys. Rev. B}\ }\textbf {\bibinfo {volume} {55}},\ \bibinfo
  {pages} {1142} (\bibinfo {year} {1997})}\BibitemShut {NoStop}%
\bibitem [{\citenamefont {Schnyder}\ \emph {et~al.}(2009)\citenamefont
  {Schnyder}, \citenamefont {Ryu}, \citenamefont {Furusaki},\ and\
  \citenamefont {Ludwig}}]{Schnyder2009}%
  \BibitemOpen
  \bibfield  {author} {\bibinfo {author} {\bibfnamefont {A.~P.}\ \bibnamefont
  {Schnyder}}, \bibinfo {author} {\bibfnamefont {S.}~\bibnamefont {Ryu}},
  \bibinfo {author} {\bibfnamefont {A.}~\bibnamefont {Furusaki}},\ and\
  \bibinfo {author} {\bibfnamefont {A.~W.~W.}\ \bibnamefont {Ludwig}},\
  }\bibfield  {title} {\bibinfo {title} {Classification of topological
  insulators and superconductors},\ }\href {https://doi.org/10.1063/1.3149481}
  {\bibfield  {journal} {\bibinfo  {journal} {AIP Conf. Proc.}\ }\textbf
  {\bibinfo {volume} {1134}},\ \bibinfo {pages} {10} (\bibinfo {year}
  {2009})}\BibitemShut {NoStop}%
\bibitem [{\citenamefont {Ryu}\ \emph {et~al.}(2010)\citenamefont {Ryu},
  \citenamefont {Schnyder}, \citenamefont {Furusaki},\ and\ \citenamefont
  {Ludwig}}]{Ryu2010}%
  \BibitemOpen
  \bibfield  {author} {\bibinfo {author} {\bibfnamefont {S.}~\bibnamefont
  {Ryu}}, \bibinfo {author} {\bibfnamefont {A.~P.}\ \bibnamefont {Schnyder}},
  \bibinfo {author} {\bibfnamefont {A.}~\bibnamefont {Furusaki}},\ and\
  \bibinfo {author} {\bibfnamefont {A.~W.~W.}\ \bibnamefont {Ludwig}},\
  }\bibfield  {title} {\bibinfo {title} {Topological insulators and
  superconductors: tenfold way and dimensional hierarchy},\ }\href
  {https://doi.org/https://doi.org/10.1088/1367-2630/12/6/065010} {\bibfield
  {journal} {\bibinfo  {journal} {New J. Phys.}\ }\textbf {\bibinfo {volume}
  {12}},\ \bibinfo {pages} {065010} (\bibinfo {year} {2010})}\BibitemShut
  {NoStop}%
\bibitem [{\citenamefont {Chiu}\ \emph {et~al.}(2016)\citenamefont {Chiu},
  \citenamefont {Teo}, \citenamefont {Schnyder},\ and\ \citenamefont
  {Ryu}}]{Chiu2016}%
  \BibitemOpen
  \bibfield  {author} {\bibinfo {author} {\bibfnamefont {C.-K.}\ \bibnamefont
  {Chiu}}, \bibinfo {author} {\bibfnamefont {J.~C.~Y.}\ \bibnamefont {Teo}},
  \bibinfo {author} {\bibfnamefont {A.~P.}\ \bibnamefont {Schnyder}},\ and\
  \bibinfo {author} {\bibfnamefont {S.}~\bibnamefont {Ryu}},\ }\bibfield
  {title} {\bibinfo {title} {Classification of topological quantum matter with
  symmetries},\ }\href {https://doi.org/10.1103/RevModPhys.88.035005}
  {\bibfield  {journal} {\bibinfo  {journal} {Rev. Mod. Phys.}\ }\textbf
  {\bibinfo {volume} {88}},\ \bibinfo {pages} {035005} (\bibinfo {year}
  {2016})}\BibitemShut {NoStop}%
\bibitem [{\citenamefont {Benalcazar}\ \emph
  {et~al.}(2017{\natexlab{a}})\citenamefont {Benalcazar}, \citenamefont
  {Bernevig},\ and\ \citenamefont {Hughes}}]{Benalcazar2017}%
  \BibitemOpen
  \bibfield  {author} {\bibinfo {author} {\bibfnamefont {W.~A.}\ \bibnamefont
  {Benalcazar}}, \bibinfo {author} {\bibfnamefont {B.~A.}\ \bibnamefont
  {Bernevig}},\ and\ \bibinfo {author} {\bibfnamefont {T.~L.}\ \bibnamefont
  {Hughes}},\ }\bibfield  {title} {\bibinfo {title} {Quantized electric
  multipole insulators},\ }\href {https://doi.org/10.1126/science.aah6442}
  {\bibfield  {journal} {\bibinfo  {journal} {Science}\ }\textbf {\bibinfo
  {volume} {357}},\ \bibinfo {pages} {61} (\bibinfo {year}
  {2017}{\natexlab{a}})}\BibitemShut {NoStop}%
\bibitem [{\citenamefont {Benalcazar}\ \emph
  {et~al.}(2017{\natexlab{b}})\citenamefont {Benalcazar}, \citenamefont
  {Bernevig},\ and\ \citenamefont {Hughes}}]{Benalcazar2017prb}%
  \BibitemOpen
  \bibfield  {author} {\bibinfo {author} {\bibfnamefont {W.~A.}\ \bibnamefont
  {Benalcazar}}, \bibinfo {author} {\bibfnamefont {B.~A.}\ \bibnamefont
  {Bernevig}},\ and\ \bibinfo {author} {\bibfnamefont {T.~L.}\ \bibnamefont
  {Hughes}},\ }\bibfield  {title} {\bibinfo {title} {Electric multipole
  moments, topological multipole moment pumping, and chiral hinge states in
  crystalline insulators},\ }\href {https://doi.org/10.1103/PhysRevB.96.245115}
  {\bibfield  {journal} {\bibinfo  {journal} {Phys. Rev. B}\ }\textbf {\bibinfo
  {volume} {96}},\ \bibinfo {pages} {245115} (\bibinfo {year}
  {2017}{\natexlab{b}})}\BibitemShut {NoStop}%
\bibitem [{\citenamefont {Langbehn}\ \emph {et~al.}(2017)\citenamefont
  {Langbehn}, \citenamefont {Peng}, \citenamefont {Trifunovic}, \citenamefont
  {von Oppen},\ and\ \citenamefont {Brouwer}}]{Langbehn2017}%
  \BibitemOpen
  \bibfield  {author} {\bibinfo {author} {\bibfnamefont {J.}~\bibnamefont
  {Langbehn}}, \bibinfo {author} {\bibfnamefont {Y.}~\bibnamefont {Peng}},
  \bibinfo {author} {\bibfnamefont {L.}~\bibnamefont {Trifunovic}}, \bibinfo
  {author} {\bibfnamefont {F.}~\bibnamefont {von Oppen}},\ and\ \bibinfo
  {author} {\bibfnamefont {P.~W.}\ \bibnamefont {Brouwer}},\ }\bibfield
  {title} {\bibinfo {title} {Reflection-symmetric second-order topological
  insulators and superconductors},\ }\href
  {https://doi.org/10.1103/PhysRevLett.119.246401} {\bibfield  {journal}
  {\bibinfo  {journal} {Phys. Rev. Lett.}\ }\textbf {\bibinfo {volume} {119}},\
  \bibinfo {pages} {246401} (\bibinfo {year} {2017})}\BibitemShut {NoStop}%
\bibitem [{\citenamefont {Song}\ \emph {et~al.}(2017)\citenamefont {Song},
  \citenamefont {Fang},\ and\ \citenamefont {Fang}}]{Song2017}%
  \BibitemOpen
  \bibfield  {author} {\bibinfo {author} {\bibfnamefont {Z.}~\bibnamefont
  {Song}}, \bibinfo {author} {\bibfnamefont {Z.}~\bibnamefont {Fang}},\ and\
  \bibinfo {author} {\bibfnamefont {C.}~\bibnamefont {Fang}},\ }\bibfield
  {title} {\bibinfo {title} {$(d\ensuremath{-}2)$-dimensional edge states of
  rotation symmetry protected topological states},\ }\href
  {https://doi.org/10.1103/PhysRevLett.119.246402} {\bibfield  {journal}
  {\bibinfo  {journal} {Phys. Rev. Lett.}\ }\textbf {\bibinfo {volume} {119}},\
  \bibinfo {pages} {246402} (\bibinfo {year} {2017})}\BibitemShut {NoStop}%
\bibitem [{\citenamefont {Schindler}\ \emph
  {et~al.}(2018{\natexlab{a}})\citenamefont {Schindler}, \citenamefont {Wang},
  \citenamefont {Vergniory}, \citenamefont {Cook}, \citenamefont {Murani},
  \citenamefont {Sengupta}, \citenamefont {Kasumov}, \citenamefont {Deblock},
  \citenamefont {Jeon}, \citenamefont {Drozdov} \emph
  {et~al.}}]{Schindler2018natphys}%
  \BibitemOpen
  \bibfield  {author} {\bibinfo {author} {\bibfnamefont {F.}~\bibnamefont
  {Schindler}}, \bibinfo {author} {\bibfnamefont {Z.}~\bibnamefont {Wang}},
  \bibinfo {author} {\bibfnamefont {M.~G.}\ \bibnamefont {Vergniory}}, \bibinfo
  {author} {\bibfnamefont {A.~M.}\ \bibnamefont {Cook}}, \bibinfo {author}
  {\bibfnamefont {A.}~\bibnamefont {Murani}}, \bibinfo {author} {\bibfnamefont
  {S.}~\bibnamefont {Sengupta}}, \bibinfo {author} {\bibfnamefont {A.~Y.}\
  \bibnamefont {Kasumov}}, \bibinfo {author} {\bibfnamefont {R.}~\bibnamefont
  {Deblock}}, \bibinfo {author} {\bibfnamefont {S.}~\bibnamefont {Jeon}},
  \bibinfo {author} {\bibfnamefont {I.}~\bibnamefont {Drozdov}}, \emph
  {et~al.},\ }\bibfield  {title} {\bibinfo {title} {Higher-order topology in
  bismuth},\ }\href {https://doi.org/10.1038/s41567-018-0224-7} {\bibfield
  {journal} {\bibinfo  {journal} {Nat. Phys.}\ }\textbf {\bibinfo {volume}
  {14}},\ \bibinfo {pages} {918} (\bibinfo {year}
  {2018}{\natexlab{a}})}\BibitemShut {NoStop}%
\bibitem [{\citenamefont {Schindler}\ \emph
  {et~al.}(2018{\natexlab{b}})\citenamefont {Schindler}, \citenamefont {Cook},
  \citenamefont {Vergniory}, \citenamefont {Wang}, \citenamefont {Parkin},
  \citenamefont {Bernevig},\ and\ \citenamefont
  {Neupert}}]{Schindler2018sciadv}%
  \BibitemOpen
  \bibfield  {author} {\bibinfo {author} {\bibfnamefont {F.}~\bibnamefont
  {Schindler}}, \bibinfo {author} {\bibfnamefont {A.~M.}\ \bibnamefont {Cook}},
  \bibinfo {author} {\bibfnamefont {M.~G.}\ \bibnamefont {Vergniory}}, \bibinfo
  {author} {\bibfnamefont {Z.}~\bibnamefont {Wang}}, \bibinfo {author}
  {\bibfnamefont {S.~S.~P.}\ \bibnamefont {Parkin}}, \bibinfo {author}
  {\bibfnamefont {B.~A.}\ \bibnamefont {Bernevig}},\ and\ \bibinfo {author}
  {\bibfnamefont {T.}~\bibnamefont {Neupert}},\ }\bibfield  {title} {\bibinfo
  {title} {Higher-order topological insulators},\ }\href
  {https://doi.org/10.1126/sciadv.aat0346} {\bibfield  {journal} {\bibinfo
  {journal} {Sci. Adv.}\ }\textbf {\bibinfo {volume} {4}},\ \bibinfo {pages}
  {0346} (\bibinfo {year} {2018}{\natexlab{b}})}\BibitemShut {NoStop}%
\bibitem [{\citenamefont {Geier}\ \emph {et~al.}(2018)\citenamefont {Geier},
  \citenamefont {Trifunovic}, \citenamefont {Hoskam},\ and\ \citenamefont
  {Brouwer}}]{Geier2018}%
  \BibitemOpen
  \bibfield  {author} {\bibinfo {author} {\bibfnamefont {M.}~\bibnamefont
  {Geier}}, \bibinfo {author} {\bibfnamefont {L.}~\bibnamefont {Trifunovic}},
  \bibinfo {author} {\bibfnamefont {M.}~\bibnamefont {Hoskam}},\ and\ \bibinfo
  {author} {\bibfnamefont {P.~W.}\ \bibnamefont {Brouwer}},\ }\bibfield
  {title} {\bibinfo {title} {Second-order topological insulators and
  superconductors with an order-two crystalline symmetry},\ }\href
  {https://doi.org/10.1103/physrevb.97.205135} {\bibfield  {journal} {\bibinfo
  {journal} {Phys. Rev. B}\ }\textbf {\bibinfo {volume} {97}},\ \bibinfo
  {pages} {205135} (\bibinfo {year} {2018})}\BibitemShut {NoStop}%
\bibitem [{\citenamefont {Khalaf}(2018)}]{Khalaf2018}%
  \BibitemOpen
  \bibfield  {author} {\bibinfo {author} {\bibfnamefont {E.}~\bibnamefont
  {Khalaf}},\ }\bibfield  {title} {\bibinfo {title} {Higher-order topological
  insulators and superconductors protected by inversion symmetry},\ }\href
  {https://doi.org/10.1103/physrevb.97.205136} {\bibfield  {journal} {\bibinfo
  {journal} {Phys. Rev. B}\ }\textbf {\bibinfo {volume} {97}},\ \bibinfo
  {pages} {205136} (\bibinfo {year} {2018})}\BibitemShut {NoStop}%
\bibitem [{\citenamefont {Trifunovic}\ and\ \citenamefont
  {Brouwer}(2019)}]{Trifunovic2019}%
  \BibitemOpen
  \bibfield  {author} {\bibinfo {author} {\bibfnamefont {L.}~\bibnamefont
  {Trifunovic}}\ and\ \bibinfo {author} {\bibfnamefont {P.~W.}\ \bibnamefont
  {Brouwer}},\ }\bibfield  {title} {\bibinfo {title} {Higher-order
  bulk-boundary correspondence for topological crystalline phases},\ }\href
  {https://doi.org/10.1103/PhysRevX.9.011012} {\bibfield  {journal} {\bibinfo
  {journal} {Phys. Rev. X}\ }\textbf {\bibinfo {volume} {9}},\ \bibinfo {pages}
  {011012} (\bibinfo {year} {2019})}\BibitemShut {NoStop}%
\bibitem [{\citenamefont {Wang}\ \emph
  {et~al.}(2018{\natexlab{a}})\citenamefont {Wang}, \citenamefont {Lin},\ and\
  \citenamefont {Hughes}}]{Hughes2018}%
  \BibitemOpen
  \bibfield  {author} {\bibinfo {author} {\bibfnamefont {Y.}~\bibnamefont
  {Wang}}, \bibinfo {author} {\bibfnamefont {M.}~\bibnamefont {Lin}},\ and\
  \bibinfo {author} {\bibfnamefont {T.~L.}\ \bibnamefont {Hughes}},\ }\bibfield
   {title} {\bibinfo {title} {Weak-pairing higher order topological
  superconductors},\ }\href {https://doi.org/10.1103/PhysRevB.98.165144}
  {\bibfield  {journal} {\bibinfo  {journal} {Phys. Rev. B}\ }\textbf {\bibinfo
  {volume} {98}},\ \bibinfo {pages} {165144} (\bibinfo {year}
  {2018}{\natexlab{a}})}\BibitemShut {NoStop}%
\bibitem [{\citenamefont {van Miert}\ and\ \citenamefont
  {Ortix}(2018)}]{Ortix2018}%
  \BibitemOpen
  \bibfield  {author} {\bibinfo {author} {\bibfnamefont {G.}~\bibnamefont {van
  Miert}}\ and\ \bibinfo {author} {\bibfnamefont {C.}~\bibnamefont {Ortix}},\
  }\bibfield  {title} {\bibinfo {title} {Higher-order topological insulators
  protected by inversion and rotoinversion symmetries},\ }\href
  {https://doi.org/10.1103/PhysRevB.98.081110} {\bibfield  {journal} {\bibinfo
  {journal} {Phys. Rev. B}\ }\textbf {\bibinfo {volume} {98}},\ \bibinfo
  {pages} {081110} (\bibinfo {year} {2018})}\BibitemShut {NoStop}%
\bibitem [{\citenamefont {You}\ \emph {et~al.}(2018)\citenamefont {You},
  \citenamefont {Devakul}, \citenamefont {Burnell},\ and\ \citenamefont
  {Neupert}}]{You2018}%
  \BibitemOpen
  \bibfield  {author} {\bibinfo {author} {\bibfnamefont {Y.}~\bibnamefont
  {You}}, \bibinfo {author} {\bibfnamefont {T.}~\bibnamefont {Devakul}},
  \bibinfo {author} {\bibfnamefont {F.~J.}\ \bibnamefont {Burnell}},\ and\
  \bibinfo {author} {\bibfnamefont {T.}~\bibnamefont {Neupert}},\ }\bibfield
  {title} {\bibinfo {title} {Higher-order symmetry-protected topological states
  for interacting bosons and fermions},\ }\href
  {https://doi.org/10.1103/PhysRevB.98.235102} {\bibfield  {journal} {\bibinfo
  {journal} {Phys. Rev. B}\ }\textbf {\bibinfo {volume} {98}},\ \bibinfo
  {pages} {235102} (\bibinfo {year} {2018})}\BibitemShut {NoStop}%
\bibitem [{\citenamefont {C\ifmmode \u{a}\else \u{a}\fi{}lug\ifmmode~\u{a}\else
  \u{a}\fi{}ru}\ \emph {et~al.}(2019)\citenamefont {C\ifmmode \u{a}\else
  \u{a}\fi{}lug\ifmmode~\u{a}\else \u{a}\fi{}ru}, \citenamefont {Juri\ifmmode
  \check{c}\else \v{c}\fi{}i\ifmmode~\acute{c}\else \'{c}\fi{}},\ and\
  \citenamefont {Roy}}]{Roy2019}%
  \BibitemOpen
  \bibfield  {author} {\bibinfo {author} {\bibfnamefont {D.}~\bibnamefont
  {C\ifmmode \u{a}\else \u{a}\fi{}lug\ifmmode~\u{a}\else \u{a}\fi{}ru}},
  \bibinfo {author} {\bibfnamefont {V.}~\bibnamefont {Juri\ifmmode
  \check{c}\else \v{c}\fi{}i\ifmmode~\acute{c}\else \'{c}\fi{}}},\ and\
  \bibinfo {author} {\bibfnamefont {B.}~\bibnamefont {Roy}},\ }\bibfield
  {title} {\bibinfo {title} {Higher-order topological phases: A general
  principle of construction},\ }\href
  {https://doi.org/10.1103/PhysRevB.99.041301} {\bibfield  {journal} {\bibinfo
  {journal} {Phys. Rev. B}\ }\textbf {\bibinfo {volume} {99}},\ \bibinfo
  {pages} {041301} (\bibinfo {year} {2019})}\BibitemShut {NoStop}%
\bibitem [{\citenamefont {Araki}\ \emph {et~al.}(2019)\citenamefont {Araki},
  \citenamefont {Mizoguchi},\ and\ \citenamefont {Hatsugai}}]{Araki2019}%
  \BibitemOpen
  \bibfield  {author} {\bibinfo {author} {\bibfnamefont {H.}~\bibnamefont
  {Araki}}, \bibinfo {author} {\bibfnamefont {T.}~\bibnamefont {Mizoguchi}},\
  and\ \bibinfo {author} {\bibfnamefont {Y.}~\bibnamefont {Hatsugai}},\
  }\bibfield  {title} {\bibinfo {title} {Phase diagram of a disordered
  higher-order topological insulator: A machine learning study},\ }\href
  {https://doi.org/10.1103/PhysRevB.99.085406} {\bibfield  {journal} {\bibinfo
  {journal} {Phys. Rev. B}\ }\textbf {\bibinfo {volume} {99}},\ \bibinfo
  {pages} {085406} (\bibinfo {year} {2019})}\BibitemShut {NoStop}%
\bibitem [{\citenamefont {Kudo}\ \emph {et~al.}(2019)\citenamefont {Kudo},
  \citenamefont {Yoshida},\ and\ \citenamefont {Hatsugai}}]{Hatsugai2019}%
  \BibitemOpen
  \bibfield  {author} {\bibinfo {author} {\bibfnamefont {K.}~\bibnamefont
  {Kudo}}, \bibinfo {author} {\bibfnamefont {T.}~\bibnamefont {Yoshida}},\ and\
  \bibinfo {author} {\bibfnamefont {Y.}~\bibnamefont {Hatsugai}},\ }\bibfield
  {title} {\bibinfo {title} {Higher-order topological {M}ott insulators},\
  }\href {https://doi.org/10.1103/PhysRevLett.123.196402} {\bibfield  {journal}
  {\bibinfo  {journal} {Phys. Rev. Lett.}\ }\textbf {\bibinfo {volume} {123}},\
  \bibinfo {pages} {196402} (\bibinfo {year} {2019})}\BibitemShut {NoStop}%
\bibitem [{\citenamefont {Tiwari}\ \emph {et~al.}(2020)\citenamefont {Tiwari},
  \citenamefont {Li}, \citenamefont {Bernevig}, \citenamefont {Neupert},\ and\
  \citenamefont {Parameswaran}}]{Tiwari2020}%
  \BibitemOpen
  \bibfield  {author} {\bibinfo {author} {\bibfnamefont {A.}~\bibnamefont
  {Tiwari}}, \bibinfo {author} {\bibfnamefont {M.-H.}\ \bibnamefont {Li}},
  \bibinfo {author} {\bibfnamefont {B.~A.}\ \bibnamefont {Bernevig}}, \bibinfo
  {author} {\bibfnamefont {T.}~\bibnamefont {Neupert}},\ and\ \bibinfo {author}
  {\bibfnamefont {S.~A.}\ \bibnamefont {Parameswaran}},\ }\bibfield  {title}
  {\bibinfo {title} {Unhinging the surfaces of higher-order topological
  insulators and superconductors},\ }\href
  {https://doi.org/10.1103/PhysRevLett.124.046801} {\bibfield  {journal}
  {\bibinfo  {journal} {Phys. Rev. Lett.}\ }\textbf {\bibinfo {volume} {124}},\
  \bibinfo {pages} {046801} (\bibinfo {year} {2020})}\BibitemShut {NoStop}%
\bibitem [{\citenamefont {Ezawa}(2018{\natexlab{a}})}]{Ezawa2018kagome}%
  \BibitemOpen
  \bibfield  {author} {\bibinfo {author} {\bibfnamefont {M.}~\bibnamefont
  {Ezawa}},\ }\bibfield  {title} {\bibinfo {title} {Higher-order topological
  insulators and semimetals on the breathing {K}agome and pyrochlore
  lattices},\ }\href {https://doi.org/10.1103/physrevlett.120.026801}
  {\bibfield  {journal} {\bibinfo  {journal} {Phys. Rev. Lett.}\ }\textbf
  {\bibinfo {volume} {120}},\ \bibinfo {pages} {026801} (\bibinfo {year}
  {2018}{\natexlab{a}})}\BibitemShut {NoStop}%
\bibitem [{\citenamefont {Ezawa}(2018{\natexlab{b}})}]{Ezawa2018phosphorene}%
  \BibitemOpen
  \bibfield  {author} {\bibinfo {author} {\bibfnamefont {M.}~\bibnamefont
  {Ezawa}},\ }\bibfield  {title} {\bibinfo {title} {Minimal models for
  {W}annier-type higher-order topological insulators and phosphorene},\ }\href
  {https://doi.org/10.1103/physrevb.98.045125} {\bibfield  {journal} {\bibinfo
  {journal} {Phys. Rev. B}\ }\textbf {\bibinfo {volume} {98}},\ \bibinfo
  {pages} {045125} (\bibinfo {year} {2018}{\natexlab{b}})}\BibitemShut
  {NoStop}%
\bibitem [{\citenamefont {Hsu}\ \emph {et~al.}(2018)\citenamefont {Hsu},
  \citenamefont {Stano}, \citenamefont {Klinovaja},\ and\ \citenamefont
  {Loss}}]{loss2018}%
  \BibitemOpen
  \bibfield  {author} {\bibinfo {author} {\bibfnamefont {C.-H.}\ \bibnamefont
  {Hsu}}, \bibinfo {author} {\bibfnamefont {P.}~\bibnamefont {Stano}}, \bibinfo
  {author} {\bibfnamefont {J.}~\bibnamefont {Klinovaja}},\ and\ \bibinfo
  {author} {\bibfnamefont {D.}~\bibnamefont {Loss}},\ }\bibfield  {title}
  {\bibinfo {title} {{M}ajorana {K}ramers pairs in higher-order topological
  insulators},\ }\href {https://doi.org/10.1103/PhysRevLett.121.196801}
  {\bibfield  {journal} {\bibinfo  {journal} {Phys. Rev. Lett.}\ }\textbf
  {\bibinfo {volume} {121}},\ \bibinfo {pages} {196801} (\bibinfo {year}
  {2018})}\BibitemShut {NoStop}%
\bibitem [{\citenamefont {Wang}\ \emph
  {et~al.}(2018{\natexlab{b}})\citenamefont {Wang}, \citenamefont {Liu},
  \citenamefont {Lu},\ and\ \citenamefont {Zhang}}]{Wang2018prl}%
  \BibitemOpen
  \bibfield  {author} {\bibinfo {author} {\bibfnamefont {Q.}~\bibnamefont
  {Wang}}, \bibinfo {author} {\bibfnamefont {C.-C.}\ \bibnamefont {Liu}},
  \bibinfo {author} {\bibfnamefont {Y.-M.}\ \bibnamefont {Lu}},\ and\ \bibinfo
  {author} {\bibfnamefont {F.}~\bibnamefont {Zhang}},\ }\bibfield  {title}
  {\bibinfo {title} {High-temperature {M}ajorana corner states},\ }\href
  {https://doi.org/10.1103/PhysRevLett.121.186801} {\bibfield  {journal}
  {\bibinfo  {journal} {Phys. Rev. Lett.}\ }\textbf {\bibinfo {volume} {121}},\
  \bibinfo {pages} {186801} (\bibinfo {year} {2018}{\natexlab{b}})}\BibitemShut
  {NoStop}%
\bibitem [{\citenamefont {Yan}\ \emph {et~al.}(2018)\citenamefont {Yan},
  \citenamefont {Song},\ and\ \citenamefont {Wang}}]{Yan2018prl}%
  \BibitemOpen
  \bibfield  {author} {\bibinfo {author} {\bibfnamefont {Z.}~\bibnamefont
  {Yan}}, \bibinfo {author} {\bibfnamefont {F.}~\bibnamefont {Song}},\ and\
  \bibinfo {author} {\bibfnamefont {Z.}~\bibnamefont {Wang}},\ }\bibfield
  {title} {\bibinfo {title} {{M}ajorana corner modes in a high-temperature
  platform},\ }\href {https://doi.org/10.1103/PhysRevLett.121.096803}
  {\bibfield  {journal} {\bibinfo  {journal} {Phys. Rev. Lett.}\ }\textbf
  {\bibinfo {volume} {121}},\ \bibinfo {pages} {096803} (\bibinfo {year}
  {2018})}\BibitemShut {NoStop}%
\bibitem [{\citenamefont {Xie}\ \emph {et~al.}(2021)\citenamefont {Xie},
  \citenamefont {Wang}, \citenamefont {Zhang}, \citenamefont {Zhan},
  \citenamefont {Jiang}, \citenamefont {Lu},\ and\ \citenamefont
  {Chen}}]{xie2021higher}%
  \BibitemOpen
  \bibfield  {author} {\bibinfo {author} {\bibfnamefont {B.}~\bibnamefont
  {Xie}}, \bibinfo {author} {\bibfnamefont {H.-X.}\ \bibnamefont {Wang}},
  \bibinfo {author} {\bibfnamefont {X.}~\bibnamefont {Zhang}}, \bibinfo
  {author} {\bibfnamefont {P.}~\bibnamefont {Zhan}}, \bibinfo {author}
  {\bibfnamefont {J.-H.}\ \bibnamefont {Jiang}}, \bibinfo {author}
  {\bibfnamefont {M.}~\bibnamefont {Lu}},\ and\ \bibinfo {author}
  {\bibfnamefont {Y.}~\bibnamefont {Chen}},\ }\bibfield  {title} {\bibinfo
  {title} {Higher-order band topology},\ }\href
  {https://doi.org/10.1038/s42254-021-00323-4} {\bibfield  {journal} {\bibinfo
  {journal} {Nat. Rev. Phys.}\ }\textbf {\bibinfo {volume} {3}},\ \bibinfo
  {pages} {520} (\bibinfo {year} {2021})}\BibitemShut {NoStop}%
\bibitem [{\citenamefont {Tokura}\ \emph {et~al.}(2019)\citenamefont {Tokura},
  \citenamefont {Yasuda},\ and\ \citenamefont {Tsukazaki}}]{Tokura2019}%
  \BibitemOpen
  \bibfield  {author} {\bibinfo {author} {\bibfnamefont {Y.}~\bibnamefont
  {Tokura}}, \bibinfo {author} {\bibfnamefont {K.}~\bibnamefont {Yasuda}},\
  and\ \bibinfo {author} {\bibfnamefont {A.}~\bibnamefont {Tsukazaki}},\
  }\bibfield  {title} {\bibinfo {title} {Magnetic topological insulators},\
  }\href {https://doi.org/10.1038/s42254-018-0011-5} {\bibfield  {journal}
  {\bibinfo  {journal} {Nature Reviews Physics}\ }\textbf {\bibinfo {volume}
  {1}},\ \bibinfo {pages} {126} (\bibinfo {year} {2019})}\BibitemShut {NoStop}%
\bibitem [{\citenamefont {Elcoro}\ \emph {et~al.}(2021)\citenamefont {Elcoro},
  \citenamefont {Wieder}, \citenamefont {Song}, \citenamefont {Xu},
  \citenamefont {Bradlyn},\ and\ \citenamefont {Bernevig}}]{Elcoro2021}%
  \BibitemOpen
  \bibfield  {author} {\bibinfo {author} {\bibfnamefont {L.}~\bibnamefont
  {Elcoro}}, \bibinfo {author} {\bibfnamefont {B.~J.}\ \bibnamefont {Wieder}},
  \bibinfo {author} {\bibfnamefont {Z.}~\bibnamefont {Song}}, \bibinfo {author}
  {\bibfnamefont {Y.}~\bibnamefont {Xu}}, \bibinfo {author} {\bibfnamefont
  {B.}~\bibnamefont {Bradlyn}},\ and\ \bibinfo {author} {\bibfnamefont {B.~A.}\
  \bibnamefont {Bernevig}},\ }\bibfield  {title} {\bibinfo {title} {Magnetic
  topological quantum chemistry},\ }\href
  {https://doi.org/10.1038/s41467-021-26241-8} {\bibfield  {journal} {\bibinfo
  {journal} {Nat. Commun.}\ }\textbf {\bibinfo {volume} {12}},\ \bibinfo
  {pages} {5965} (\bibinfo {year} {2021})}\BibitemShut {NoStop}%
\bibitem [{\citenamefont {Bernevig}\ \emph {et~al.}(2022)\citenamefont
  {Bernevig}, \citenamefont {Felser},\ and\ \citenamefont
  {Beidenkopf}}]{Bernevig2022}%
  \BibitemOpen
  \bibfield  {author} {\bibinfo {author} {\bibfnamefont {B.~A.}\ \bibnamefont
  {Bernevig}}, \bibinfo {author} {\bibfnamefont {C.}~\bibnamefont {Felser}},\
  and\ \bibinfo {author} {\bibfnamefont {H.}~\bibnamefont {Beidenkopf}},\
  }\bibfield  {title} {\bibinfo {title} {Progress and prospects in magnetic
  topological materials},\ }\href {https://doi.org/10.1038/s41586-021-04105-x}
  {\bibfield  {journal} {\bibinfo  {journal} {Nature}\ }\textbf {\bibinfo
  {volume} {603}},\ \bibinfo {pages} {41} (\bibinfo {year} {2022})}\BibitemShut
  {NoStop}%
\bibitem [{\citenamefont {Mong}\ \emph {et~al.}(2010)\citenamefont {Mong},
  \citenamefont {Essin},\ and\ \citenamefont {Moore}}]{ati}%
  \BibitemOpen
  \bibfield  {author} {\bibinfo {author} {\bibfnamefont {R.~S.~K.}\
  \bibnamefont {Mong}}, \bibinfo {author} {\bibfnamefont {A.~M.}\ \bibnamefont
  {Essin}},\ and\ \bibinfo {author} {\bibfnamefont {J.~E.}\ \bibnamefont
  {Moore}},\ }\bibfield  {title} {\bibinfo {title} {Antiferromagnetic
  topological insulators},\ }\href {https://doi.org/10.1103/PhysRevB.81.245209}
  {\bibfield  {journal} {\bibinfo  {journal} {Phys. Rev. B}\ }\textbf {\bibinfo
  {volume} {81}},\ \bibinfo {pages} {245209} (\bibinfo {year}
  {2010})}\BibitemShut {NoStop}%
\bibitem [{\citenamefont {Otrokov}\ \emph {et~al.}(2019)\citenamefont
  {Otrokov}, \citenamefont {Klimovskikh}, \citenamefont {Bentmann},
  \citenamefont {Estyunin}, \citenamefont {Zeugner}, \citenamefont {Aliev},
  \citenamefont {Gaß}, \citenamefont {Wolter}, \citenamefont {Koroleva},
  \citenamefont {Shikin}, \citenamefont {Blanco-Rey}, \citenamefont {Hoffmann},
  \citenamefont {Rusinov}, \citenamefont {Vyazovskaya}, \citenamefont
  {Eremeev}, \citenamefont {Koroteev}, \citenamefont {Kuznetsov}, \citenamefont
  {Freyse}, \citenamefont {Sánchez-Barriga}, \citenamefont {Amiraslanov},
  \citenamefont {Babanly}, \citenamefont {Mamedov}, \citenamefont {Abdullayev},
  \citenamefont {Zverev}, \citenamefont {Alfonsov}, \citenamefont {Kataev},
  \citenamefont {Büchner}, \citenamefont {Schwier}, \citenamefont {Kumar},
  \citenamefont {Kimura}, \citenamefont {Petaccia}, \citenamefont {Di~Santo},
  \citenamefont {Vidal}, \citenamefont {Schatz}, \citenamefont {Kißner},
  \citenamefont {Ünzelmann}, \citenamefont {Min}, \citenamefont {Moser},
  \citenamefont {Peixoto}, \citenamefont {Reinert}, \citenamefont {Ernst},
  \citenamefont {Echenique}, \citenamefont {Isaeva},\ and\ \citenamefont
  {Chulkov}}]{Otrokov2019}%
  \BibitemOpen
  \bibfield  {author} {\bibinfo {author} {\bibfnamefont {M.~M.}\ \bibnamefont
  {Otrokov}}, \bibinfo {author} {\bibfnamefont {I.~I.}\ \bibnamefont
  {Klimovskikh}}, \bibinfo {author} {\bibfnamefont {H.}~\bibnamefont
  {Bentmann}}, \bibinfo {author} {\bibfnamefont {D.}~\bibnamefont {Estyunin}},
  \bibinfo {author} {\bibfnamefont {A.}~\bibnamefont {Zeugner}}, \bibinfo
  {author} {\bibfnamefont {Z.~S.}\ \bibnamefont {Aliev}}, \bibinfo {author}
  {\bibfnamefont {S.}~\bibnamefont {Gaß}}, \bibinfo {author} {\bibfnamefont
  {A.~U.~B.}\ \bibnamefont {Wolter}}, \bibinfo {author} {\bibfnamefont {A.~V.}\
  \bibnamefont {Koroleva}}, \bibinfo {author} {\bibfnamefont {A.~M.}\
  \bibnamefont {Shikin}}, \bibinfo {author} {\bibfnamefont {M.}~\bibnamefont
  {Blanco-Rey}}, \bibinfo {author} {\bibfnamefont {M.}~\bibnamefont
  {Hoffmann}}, \bibinfo {author} {\bibfnamefont {I.~P.}\ \bibnamefont
  {Rusinov}}, \bibinfo {author} {\bibfnamefont {A.~Y.}\ \bibnamefont
  {Vyazovskaya}}, \bibinfo {author} {\bibfnamefont {S.~V.}\ \bibnamefont
  {Eremeev}}, \bibinfo {author} {\bibfnamefont {Y.~M.}\ \bibnamefont
  {Koroteev}}, \bibinfo {author} {\bibfnamefont {V.~M.}\ \bibnamefont
  {Kuznetsov}}, \bibinfo {author} {\bibfnamefont {F.}~\bibnamefont {Freyse}},
  \bibinfo {author} {\bibfnamefont {J.}~\bibnamefont {Sánchez-Barriga}},
  \bibinfo {author} {\bibfnamefont {I.~R.}\ \bibnamefont {Amiraslanov}},
  \bibinfo {author} {\bibfnamefont {M.~B.}\ \bibnamefont {Babanly}}, \bibinfo
  {author} {\bibfnamefont {N.~T.}\ \bibnamefont {Mamedov}}, \bibinfo {author}
  {\bibfnamefont {N.~A.}\ \bibnamefont {Abdullayev}}, \bibinfo {author}
  {\bibfnamefont {V.~N.}\ \bibnamefont {Zverev}}, \bibinfo {author}
  {\bibfnamefont {A.}~\bibnamefont {Alfonsov}}, \bibinfo {author}
  {\bibfnamefont {V.}~\bibnamefont {Kataev}}, \bibinfo {author} {\bibfnamefont
  {B.}~\bibnamefont {Büchner}}, \bibinfo {author} {\bibfnamefont {E.~F.}\
  \bibnamefont {Schwier}}, \bibinfo {author} {\bibfnamefont {S.}~\bibnamefont
  {Kumar}}, \bibinfo {author} {\bibfnamefont {A.}~\bibnamefont {Kimura}},
  \bibinfo {author} {\bibfnamefont {L.}~\bibnamefont {Petaccia}}, \bibinfo
  {author} {\bibfnamefont {G.}~\bibnamefont {Di~Santo}}, \bibinfo {author}
  {\bibfnamefont {R.~C.}\ \bibnamefont {Vidal}}, \bibinfo {author}
  {\bibfnamefont {S.}~\bibnamefont {Schatz}}, \bibinfo {author} {\bibfnamefont
  {K.}~\bibnamefont {Kißner}}, \bibinfo {author} {\bibfnamefont
  {M.}~\bibnamefont {Ünzelmann}}, \bibinfo {author} {\bibfnamefont {C.~H.}\
  \bibnamefont {Min}}, \bibinfo {author} {\bibfnamefont {S.}~\bibnamefont
  {Moser}}, \bibinfo {author} {\bibfnamefont {T.~R.~F.}\ \bibnamefont
  {Peixoto}}, \bibinfo {author} {\bibfnamefont {F.}~\bibnamefont {Reinert}},
  \bibinfo {author} {\bibfnamefont {A.}~\bibnamefont {Ernst}}, \bibinfo
  {author} {\bibfnamefont {P.~M.}\ \bibnamefont {Echenique}}, \bibinfo {author}
  {\bibfnamefont {A.}~\bibnamefont {Isaeva}},\ and\ \bibinfo {author}
  {\bibfnamefont {E.~V.}\ \bibnamefont {Chulkov}},\ }\bibfield  {title}
  {\bibinfo {title} {Prediction and observation of an antiferromagnetic
  topological insulator},\ }\href {https://doi.org/10.1038/s41586-019-1840-9}
  {\bibfield  {journal} {\bibinfo  {journal} {Nature}\ }\textbf {\bibinfo
  {volume} {576}},\ \bibinfo {pages} {416} (\bibinfo {year}
  {2019})}\BibitemShut {NoStop}%
\bibitem [{\citenamefont {\ifmmode~\check{S}\else \v{S}\fi{}mejkal}\ \emph
  {et~al.}(2022{\natexlab{a}})\citenamefont {\ifmmode~\check{S}\else
  \v{S}\fi{}mejkal}, \citenamefont {Sinova},\ and\ \citenamefont
  {Jungwirth}}]{altermagnetism1}%
  \BibitemOpen
  \bibfield  {author} {\bibinfo {author} {\bibfnamefont {L.}~\bibnamefont
  {\ifmmode~\check{S}\else \v{S}\fi{}mejkal}}, \bibinfo {author} {\bibfnamefont
  {J.}~\bibnamefont {Sinova}},\ and\ \bibinfo {author} {\bibfnamefont
  {T.}~\bibnamefont {Jungwirth}},\ }\bibfield  {title} {\bibinfo {title}
  {Emerging research landscape of altermagnetism},\ }\href
  {https://doi.org/10.1103/PhysRevX.12.040501} {\bibfield  {journal} {\bibinfo
  {journal} {Phys. Rev. X}\ }\textbf {\bibinfo {volume} {12}},\ \bibinfo
  {pages} {040501} (\bibinfo {year} {2022}{\natexlab{a}})}\BibitemShut
  {NoStop}%
\bibitem [{\citenamefont {\ifmmode~\check{S}\else \v{S}\fi{}mejkal}\ \emph
  {et~al.}(2022{\natexlab{b}})\citenamefont {\ifmmode~\check{S}\else
  \v{S}\fi{}mejkal}, \citenamefont {Sinova},\ and\ \citenamefont
  {Jungwirth}}]{altermagnetism2}%
  \BibitemOpen
  \bibfield  {author} {\bibinfo {author} {\bibfnamefont {L.}~\bibnamefont
  {\ifmmode~\check{S}\else \v{S}\fi{}mejkal}}, \bibinfo {author} {\bibfnamefont
  {J.}~\bibnamefont {Sinova}},\ and\ \bibinfo {author} {\bibfnamefont
  {T.}~\bibnamefont {Jungwirth}},\ }\bibfield  {title} {\bibinfo {title}
  {Beyond conventional ferromagnetism and antiferromagnetism: A phase with
  nonrelativistic spin and crystal rotation symmetry},\ }\href
  {https://doi.org/10.1103/PhysRevX.12.031042} {\bibfield  {journal} {\bibinfo
  {journal} {Phys. Rev. X}\ }\textbf {\bibinfo {volume} {12}},\ \bibinfo
  {pages} {031042} (\bibinfo {year} {2022}{\natexlab{b}})}\BibitemShut
  {NoStop}%
\bibitem [{\citenamefont {Guo}\ \emph {et~al.}(2023)\citenamefont {Guo},
  \citenamefont {Liu}, \citenamefont {Janson}, \citenamefont {Fulga},
  \citenamefont {{van den Brink}},\ and\ \citenamefont {Facio}}]{Guo2023}%
  \BibitemOpen
  \bibfield  {author} {\bibinfo {author} {\bibfnamefont {Y.}~\bibnamefont
  {Guo}}, \bibinfo {author} {\bibfnamefont {H.}~\bibnamefont {Liu}}, \bibinfo
  {author} {\bibfnamefont {O.}~\bibnamefont {Janson}}, \bibinfo {author}
  {\bibfnamefont {I.~C.}\ \bibnamefont {Fulga}}, \bibinfo {author}
  {\bibfnamefont {J.}~\bibnamefont {{van den Brink}}},\ and\ \bibinfo {author}
  {\bibfnamefont {J.~I.}\ \bibnamefont {Facio}},\ }\bibfield  {title} {\bibinfo
  {title} {Spin-split collinear antiferromagnets: A large-scale ab-initio
  study},\ }\href
  {https://doi.org/https://doi.org/10.1016/j.mtphys.2023.100991} {\bibfield
  {journal} {\bibinfo  {journal} {Materials Today Physics}\ }\textbf {\bibinfo
  {volume} {32}},\ \bibinfo {pages} {100991} (\bibinfo {year}
  {2023})}\BibitemShut {NoStop}%
\bibitem [{\citenamefont {Chalker}\ and\ \citenamefont
  {Coddington}(1988)}]{chalker1988}%
  \BibitemOpen
  \bibfield  {author} {\bibinfo {author} {\bibfnamefont {J.~T.}\ \bibnamefont
  {Chalker}}\ and\ \bibinfo {author} {\bibfnamefont {P.~D.}\ \bibnamefont
  {Coddington}},\ }\bibfield  {title} {\bibinfo {title} {Percolation, quantum
  tunnelling and the integer {H}all effect},\ }\href
  {https://doi.org/10.1088/0022-3719/21/14/008} {\bibfield  {journal} {\bibinfo
   {journal} {J. Phys. C: Solid State Physics}\ }\textbf {\bibinfo {volume}
  {21}},\ \bibinfo {pages} {2665} (\bibinfo {year} {1988})}\BibitemShut
  {NoStop}%
\bibitem [{\citenamefont {Kramer}\ \emph {et~al.}(2005)\citenamefont {Kramer},
  \citenamefont {Ohtsuki},\ and\ \citenamefont {Kettemann}}]{kramer2005review}%
  \BibitemOpen
  \bibfield  {author} {\bibinfo {author} {\bibfnamefont {B.}~\bibnamefont
  {Kramer}}, \bibinfo {author} {\bibfnamefont {T.}~\bibnamefont {Ohtsuki}},\
  and\ \bibinfo {author} {\bibfnamefont {S.}~\bibnamefont {Kettemann}},\
  }\bibfield  {title} {\bibinfo {title} {Random network models and quantum
  phase transitions in two dimensions},\ }\href
  {https://doi.org/https://doi.org/10.1016/j.physrep.2005.07.001} {\bibfield
  {journal} {\bibinfo  {journal} {Phys. Rep.}\ }\textbf {\bibinfo {volume}
  {417}},\ \bibinfo {pages} {211} (\bibinfo {year} {2005})}\BibitemShut
  {NoStop}%
\bibitem [{\citenamefont {Kagalovsky}\ \emph {et~al.}(1999)\citenamefont
  {Kagalovsky}, \citenamefont {Horovitz}, \citenamefont {Avishai},\ and\
  \citenamefont {Chalker}}]{chalker1999super}%
  \BibitemOpen
  \bibfield  {author} {\bibinfo {author} {\bibfnamefont {V.}~\bibnamefont
  {Kagalovsky}}, \bibinfo {author} {\bibfnamefont {B.}~\bibnamefont
  {Horovitz}}, \bibinfo {author} {\bibfnamefont {Y.}~\bibnamefont {Avishai}},\
  and\ \bibinfo {author} {\bibfnamefont {J.~T.}\ \bibnamefont {Chalker}},\
  }\bibfield  {title} {\bibinfo {title} {Quantum {H}all plateau transitions in
  disordered superconductors},\ }\href
  {https://doi.org/10.1103/PhysRevLett.82.3516} {\bibfield  {journal} {\bibinfo
   {journal} {Phys. Rev. Lett.}\ }\textbf {\bibinfo {volume} {82}},\ \bibinfo
  {pages} {3516} (\bibinfo {year} {1999})}\BibitemShut {NoStop}%
\bibitem [{\citenamefont {Chalker}\ \emph {et~al.}(2001)\citenamefont
  {Chalker}, \citenamefont {Read}, \citenamefont {Kagalovsky}, \citenamefont
  {Horovitz}, \citenamefont {Avishai},\ and\ \citenamefont
  {Ludwig}}]{chalker2001super}%
  \BibitemOpen
  \bibfield  {author} {\bibinfo {author} {\bibfnamefont {J.~T.}\ \bibnamefont
  {Chalker}}, \bibinfo {author} {\bibfnamefont {N.}~\bibnamefont {Read}},
  \bibinfo {author} {\bibfnamefont {V.}~\bibnamefont {Kagalovsky}}, \bibinfo
  {author} {\bibfnamefont {B.}~\bibnamefont {Horovitz}}, \bibinfo {author}
  {\bibfnamefont {Y.}~\bibnamefont {Avishai}},\ and\ \bibinfo {author}
  {\bibfnamefont {A.~W.~W.}\ \bibnamefont {Ludwig}},\ }\bibfield  {title}
  {\bibinfo {title} {Thermal metal in network models of a disordered
  two-dimensional superconductor},\ }\href
  {https://doi.org/10.1103/PhysRevB.65.012506} {\bibfield  {journal} {\bibinfo
  {journal} {Phys. Rev. B}\ }\textbf {\bibinfo {volume} {65}},\ \bibinfo
  {pages} {012506} (\bibinfo {year} {2001})}\BibitemShut {NoStop}%
\bibitem [{\citenamefont {Beamond}\ \emph {et~al.}(2002)\citenamefont
  {Beamond}, \citenamefont {Cardy},\ and\ \citenamefont
  {Chalker}}]{chalker2002qshe}%
  \BibitemOpen
  \bibfield  {author} {\bibinfo {author} {\bibfnamefont {E.~J.}\ \bibnamefont
  {Beamond}}, \bibinfo {author} {\bibfnamefont {J.}~\bibnamefont {Cardy}},\
  and\ \bibinfo {author} {\bibfnamefont {J.~T.}\ \bibnamefont {Chalker}},\
  }\bibfield  {title} {\bibinfo {title} {Quantum and classical localization,
  the spin quantum {H}all effect, and generalizations},\ }\href
  {https://doi.org/10.1103/PhysRevB.65.214301} {\bibfield  {journal} {\bibinfo
  {journal} {Phys. Rev. B}\ }\textbf {\bibinfo {volume} {65}},\ \bibinfo
  {pages} {214301} (\bibinfo {year} {2002})}\BibitemShut {NoStop}%
\bibitem [{\citenamefont {Obuse}\ \emph {et~al.}(2007)\citenamefont {Obuse},
  \citenamefont {Furusaki}, \citenamefont {Ryu},\ and\ \citenamefont
  {Mudry}}]{murdy2007qshe}%
  \BibitemOpen
  \bibfield  {author} {\bibinfo {author} {\bibfnamefont {H.}~\bibnamefont
  {Obuse}}, \bibinfo {author} {\bibfnamefont {A.}~\bibnamefont {Furusaki}},
  \bibinfo {author} {\bibfnamefont {S.}~\bibnamefont {Ryu}},\ and\ \bibinfo
  {author} {\bibfnamefont {C.}~\bibnamefont {Mudry}},\ }\bibfield  {title}
  {\bibinfo {title} {Two-dimensional spin-filtered chiral network model for the
  $\mathbbm{Z}_{2}$ quantum spin-{H}all effect},\ }\href
  {https://doi.org/10.1103/PhysRevB.76.075301} {\bibfield  {journal} {\bibinfo
  {journal} {Phys. Rev. B}\ }\textbf {\bibinfo {volume} {76}},\ \bibinfo
  {pages} {075301} (\bibinfo {year} {2007})}\BibitemShut {NoStop}%
\bibitem [{\citenamefont {Obuse}\ \emph {et~al.}(2014)\citenamefont {Obuse},
  \citenamefont {Ryu}, \citenamefont {Furusaki},\ and\ \citenamefont
  {Mudry}}]{murdy2014TI}%
  \BibitemOpen
  \bibfield  {author} {\bibinfo {author} {\bibfnamefont {H.}~\bibnamefont
  {Obuse}}, \bibinfo {author} {\bibfnamefont {S.}~\bibnamefont {Ryu}}, \bibinfo
  {author} {\bibfnamefont {A.}~\bibnamefont {Furusaki}},\ and\ \bibinfo
  {author} {\bibfnamefont {C.}~\bibnamefont {Mudry}},\ }\bibfield  {title}
  {\bibinfo {title} {Spin-directed network model for the surface states of weak
  three-dimensional $\mathbbm{Z}_{2}$ topological insulators},\ }\href
  {https://doi.org/10.1103/PhysRevB.89.155315} {\bibfield  {journal} {\bibinfo
  {journal} {Phys. Rev. B}\ }\textbf {\bibinfo {volume} {89}},\ \bibinfo
  {pages} {155315} (\bibinfo {year} {2014})}\BibitemShut {NoStop}%
\bibitem [{\citenamefont {Fulga}\ \emph
  {et~al.}(2012{\natexlab{a}})\citenamefont {Fulga}, \citenamefont {Akhmerov},
  \citenamefont {Tworzyd\l{}o}, \citenamefont {B\'eri},\ and\ \citenamefont
  {Beenakker}}]{fulga2012thermal}%
  \BibitemOpen
  \bibfield  {author} {\bibinfo {author} {\bibfnamefont {I.~C.}\ \bibnamefont
  {Fulga}}, \bibinfo {author} {\bibfnamefont {A.~R.}\ \bibnamefont {Akhmerov}},
  \bibinfo {author} {\bibfnamefont {J.}~\bibnamefont {Tworzyd\l{}o}}, \bibinfo
  {author} {\bibfnamefont {B.}~\bibnamefont {B\'eri}},\ and\ \bibinfo {author}
  {\bibfnamefont {C.~W.~J.}\ \bibnamefont {Beenakker}},\ }\bibfield  {title}
  {\bibinfo {title} {Thermal metal-insulator transition in a helical
  topological superconductor},\ }\href
  {https://doi.org/10.1103/PhysRevB.86.054505} {\bibfield  {journal} {\bibinfo
  {journal} {Phys. Rev. B}\ }\textbf {\bibinfo {volume} {86}},\ \bibinfo
  {pages} {054505} (\bibinfo {year} {2012}{\natexlab{a}})}\BibitemShut
  {NoStop}%
\bibitem [{\citenamefont {Liu}\ \emph {et~al.}(2021)\citenamefont {Liu},
  \citenamefont {Franca}, \citenamefont {Moghaddam}, \citenamefont {Hassler},\
  and\ \citenamefont {Fulga}}]{Liu2020-hoti}%
  \BibitemOpen
  \bibfield  {author} {\bibinfo {author} {\bibfnamefont {H.}~\bibnamefont
  {Liu}}, \bibinfo {author} {\bibfnamefont {S.}~\bibnamefont {Franca}},
  \bibinfo {author} {\bibfnamefont {A.~G.}\ \bibnamefont {Moghaddam}}, \bibinfo
  {author} {\bibfnamefont {F.}~\bibnamefont {Hassler}},\ and\ \bibinfo {author}
  {\bibfnamefont {I.~C.}\ \bibnamefont {Fulga}},\ }\bibfield  {title} {\bibinfo
  {title} {Network model for higher-order topological phases},\ }\href
  {https://doi.org/10.1103/PhysRevB.103.115428} {\bibfield  {journal} {\bibinfo
   {journal} {Phys. Rev. B}\ }\textbf {\bibinfo {volume} {103}},\ \bibinfo
  {pages} {115428} (\bibinfo {year} {2021})}\BibitemShut {NoStop}%
\bibitem [{\citenamefont {Ho}\ and\ \citenamefont {Chalker}(1996)}]{ho1996}%
  \BibitemOpen
  \bibfield  {author} {\bibinfo {author} {\bibfnamefont {C.-M.}\ \bibnamefont
  {Ho}}\ and\ \bibinfo {author} {\bibfnamefont {J.~T.}\ \bibnamefont
  {Chalker}},\ }\bibfield  {title} {\bibinfo {title} {Models for the integer
  quantum {H}all effect: The network model, the {D}irac equation, and a
  tight-binding {H}amiltonian},\ }\href
  {https://doi.org/10.1103/PhysRevB.54.8708} {\bibfield  {journal} {\bibinfo
  {journal} {Phys. Rev. B}\ }\textbf {\bibinfo {volume} {54}},\ \bibinfo
  {pages} {8708} (\bibinfo {year} {1996})}\BibitemShut {NoStop}%
\bibitem [{\citenamefont {Janssen}\ \emph {et~al.}(1999)\citenamefont
  {Janssen}, \citenamefont {Metzler},\ and\ \citenamefont
  {Zirnbauer}}]{zirnbauer1999}%
  \BibitemOpen
  \bibfield  {author} {\bibinfo {author} {\bibfnamefont {M.}~\bibnamefont
  {Janssen}}, \bibinfo {author} {\bibfnamefont {M.}~\bibnamefont {Metzler}},\
  and\ \bibinfo {author} {\bibfnamefont {M.~R.}\ \bibnamefont {Zirnbauer}},\
  }\bibfield  {title} {\bibinfo {title} {Point-contact conductances at the
  quantum {H}all transition},\ }\href
  {https://doi.org/10.1103/PhysRevB.59.15836} {\bibfield  {journal} {\bibinfo
  {journal} {Phys. Rev. B}\ }\textbf {\bibinfo {volume} {59}},\ \bibinfo
  {pages} {15836} (\bibinfo {year} {1999})}\BibitemShut {NoStop}%
\bibitem [{\citenamefont {Potter}\ \emph {et~al.}(2020)\citenamefont {Potter},
  \citenamefont {Chalker},\ and\ \citenamefont {Gurarie}}]{potter2020}%
  \BibitemOpen
  \bibfield  {author} {\bibinfo {author} {\bibfnamefont {A.~C.}\ \bibnamefont
  {Potter}}, \bibinfo {author} {\bibfnamefont {J.~T.}\ \bibnamefont
  {Chalker}},\ and\ \bibinfo {author} {\bibfnamefont {V.}~\bibnamefont
  {Gurarie}},\ }\bibfield  {title} {\bibinfo {title} {Quantum hall network
  models as floquet topological insulators},\ }\href
  {https://doi.org/10.1103/PhysRevLett.125.086601} {\bibfield  {journal}
  {\bibinfo  {journal} {Phys. Rev. Lett.}\ }\textbf {\bibinfo {volume} {125}},\
  \bibinfo {pages} {086601} (\bibinfo {year} {2020})}\BibitemShut {NoStop}%
\bibitem [{\citenamefont {Delplace}(2020)}]{Phase_rotation1}%
  \BibitemOpen
  \bibfield  {author} {\bibinfo {author} {\bibfnamefont {P.~A.~L.}\
  \bibnamefont {Delplace}},\ }\bibfield  {title} {\bibinfo {title}
  {{Topological chiral modes in random scattering networks}},\ }\href
  {https://doi.org/10.21468/SciPostPhys.8.5.081} {\bibfield  {journal}
  {\bibinfo  {journal} {SciPost Phys.}\ }\textbf {\bibinfo {volume} {8}},\
  \bibinfo {pages} {081} (\bibinfo {year} {2020})}\BibitemShut {NoStop}%
\bibitem [{\citenamefont {Delplace}\ \emph {et~al.}(2017)\citenamefont
  {Delplace}, \citenamefont {Fruchart},\ and\ \citenamefont
  {Tauber}}]{Phase_rotation2}%
  \BibitemOpen
  \bibfield  {author} {\bibinfo {author} {\bibfnamefont {P.}~\bibnamefont
  {Delplace}}, \bibinfo {author} {\bibfnamefont {M.}~\bibnamefont {Fruchart}},\
  and\ \bibinfo {author} {\bibfnamefont {C.}~\bibnamefont {Tauber}},\
  }\bibfield  {title} {\bibinfo {title} {Phase rotation symmetry and the
  topology of oriented scattering networks},\ }\href
  {https://doi.org/10.1103/PhysRevB.95.205413} {\bibfield  {journal} {\bibinfo
  {journal} {Phys. Rev. B}\ }\textbf {\bibinfo {volume} {95}},\ \bibinfo
  {pages} {205413} (\bibinfo {year} {2017})}\BibitemShut {NoStop}%
\bibitem [{\citenamefont {{Araya Day}}\ \emph {et~al.}(2023)\citenamefont
  {{Araya Day}}, \citenamefont {Varentcova}, \citenamefont {Varjas},\ and\
  \citenamefont {Akhmerov}}]{pfaffian}%
  \BibitemOpen
  \bibfield  {author} {\bibinfo {author} {\bibfnamefont {I.}~\bibnamefont
  {{Araya Day}}}, \bibinfo {author} {\bibfnamefont {A.}~\bibnamefont
  {Varentcova}}, \bibinfo {author} {\bibfnamefont {D.}~\bibnamefont {Varjas}},\
  and\ \bibinfo {author} {\bibfnamefont {A.~R.}\ \bibnamefont {Akhmerov}},\
  }\bibfield  {title} {\bibinfo {title} {{Pfaffian invariant identifies
  magnetic obstructed atomic insulators}},\ }\href
  {https://doi.org/10.21468/SciPostPhys.15.3.114} {\bibfield  {journal}
  {\bibinfo  {journal} {SciPost Phys.}\ }\textbf {\bibinfo {volume} {15}},\
  \bibinfo {pages} {114} (\bibinfo {year} {2023})}\BibitemShut {NoStop}%
\bibitem [{\citenamefont {Beenakker}(1997)}]{Beenakker_review}%
  \BibitemOpen
  \bibfield  {author} {\bibinfo {author} {\bibfnamefont {C.~W.~J.}\
  \bibnamefont {Beenakker}},\ }\bibfield  {title} {\bibinfo {title}
  {Random-matrix theory of quantum transport},\ }\href
  {https://doi.org/10.1103/RevModPhys.69.731} {\bibfield  {journal} {\bibinfo
  {journal} {Rev. Mod. Phys.}\ }\textbf {\bibinfo {volume} {69}},\ \bibinfo
  {pages} {731} (\bibinfo {year} {1997})}\BibitemShut {NoStop}%
\bibitem [{\citenamefont {Trifunovic}\ and\ \citenamefont
  {Brouwer}(2021)}]{Trifunovic_intrinsic}%
  \BibitemOpen
  \bibfield  {author} {\bibinfo {author} {\bibfnamefont {L.}~\bibnamefont
  {Trifunovic}}\ and\ \bibinfo {author} {\bibfnamefont {P.~W.}\ \bibnamefont
  {Brouwer}},\ }\bibfield  {title} {\bibinfo {title} {Higher-order topological
  band structures},\ }\href
  {https://doi.org/https://doi.org/10.1002/pssb.202000090} {\bibfield
  {journal} {\bibinfo  {journal} {physica status solidi (b)}\ }\textbf
  {\bibinfo {volume} {258}},\ \bibinfo {pages} {2000090} (\bibinfo {year}
  {2021})}\BibitemShut {NoStop}%
\bibitem [{\citenamefont {Su}\ \emph {et~al.}(1979)\citenamefont {Su},
  \citenamefont {Schrieffer},\ and\ \citenamefont {Heeger}}]{ssh}%
  \BibitemOpen
  \bibfield  {author} {\bibinfo {author} {\bibfnamefont {W.~P.}\ \bibnamefont
  {Su}}, \bibinfo {author} {\bibfnamefont {J.~R.}\ \bibnamefont {Schrieffer}},\
  and\ \bibinfo {author} {\bibfnamefont {A.~J.}\ \bibnamefont {Heeger}},\
  }\bibfield  {title} {\bibinfo {title} {Solitons in polyacetylene},\ }\href
  {https://doi.org/10.1103/PhysRevLett.42.1698} {\bibfield  {journal} {\bibinfo
   {journal} {Phys. Rev. Lett.}\ }\textbf {\bibinfo {volume} {42}},\ \bibinfo
  {pages} {1698} (\bibinfo {year} {1979})}\BibitemShut {NoStop}%
\bibitem [{Note1()}]{Note1}%
  \BibitemOpen
  \bibinfo {note} {P. W. Brouwer, Ph.D. thesis, Leiden University,
  1997.}\BibitemShut {Stop}%
\bibitem [{\citenamefont {Fulga}\ \emph
  {et~al.}(2012{\natexlab{b}})\citenamefont {Fulga}, \citenamefont {Hassler},\
  and\ \citenamefont {Akhmerov}}]{scattering_invariant}%
  \BibitemOpen
  \bibfield  {author} {\bibinfo {author} {\bibfnamefont {I.~C.}\ \bibnamefont
  {Fulga}}, \bibinfo {author} {\bibfnamefont {F.}~\bibnamefont {Hassler}},\
  and\ \bibinfo {author} {\bibfnamefont {A.~R.}\ \bibnamefont {Akhmerov}},\
  }\bibfield  {title} {\bibinfo {title} {Scattering theory of topological
  insulators and superconductors},\ }\href
  {https://doi.org/10.1103/PhysRevB.85.165409} {\bibfield  {journal} {\bibinfo
  {journal} {Phys. Rev. B}\ }\textbf {\bibinfo {volume} {85}},\ \bibinfo
  {pages} {165409} (\bibinfo {year} {2012}{\natexlab{b}})}\BibitemShut
  {NoStop}%
\bibitem [{\citenamefont {Po}(2020)}]{Po2020}%
  \BibitemOpen
  \bibfield  {author} {\bibinfo {author} {\bibfnamefont {H.~C.}\ \bibnamefont
  {Po}},\ }\bibfield  {title} {\bibinfo {title} {Symmetry indicators of band
  topology},\ }\href {https://doi.org/10.1088/1361-648X/ab7adb} {\bibfield
  {journal} {\bibinfo  {journal} {J. Phys. Condens. Matter}\ }\textbf {\bibinfo
  {volume} {32}},\ \bibinfo {pages} {263001} (\bibinfo {year}
  {2020})}\BibitemShut {NoStop}%
\bibitem [{\citenamefont {Po}\ \emph {et~al.}(2017)\citenamefont {Po},
  \citenamefont {Vishwanath},\ and\ \citenamefont {Watanabe}}]{Po2017a}%
  \BibitemOpen
  \bibfield  {author} {\bibinfo {author} {\bibfnamefont {H.~C.}\ \bibnamefont
  {Po}}, \bibinfo {author} {\bibfnamefont {A.}~\bibnamefont {Vishwanath}},\
  and\ \bibinfo {author} {\bibfnamefont {H.}~\bibnamefont {Watanabe}},\
  }\bibfield  {title} {\bibinfo {title} {Symmetry-based indicators of band
  topology in the 230 space groups},\ }\href
  {https://doi.org/10.1038/s41467-017-00133-2} {\bibfield  {journal} {\bibinfo
  {journal} {Nat. Commun.}\ }\textbf {\bibinfo {volume} {8}},\ \bibinfo {pages}
  {50} (\bibinfo {year} {2017})}\BibitemShut {NoStop}%
\bibitem [{\citenamefont {Kane}\ and\ \citenamefont {Mele}(2005)}]{qsh2}%
  \BibitemOpen
  \bibfield  {author} {\bibinfo {author} {\bibfnamefont {C.~L.}\ \bibnamefont
  {Kane}}\ and\ \bibinfo {author} {\bibfnamefont {E.~J.}\ \bibnamefont
  {Mele}},\ }\bibfield  {title} {\bibinfo {title} {Quantum spin hall effect in
  graphene},\ }\href {https://doi.org/10.1103/PhysRevLett.95.226801} {\bibfield
   {journal} {\bibinfo  {journal} {Phys. Rev. Lett.}\ }\textbf {\bibinfo
  {volume} {95}},\ \bibinfo {pages} {226801} (\bibinfo {year}
  {2005})}\BibitemShut {NoStop}%
\bibitem [{\citenamefont {Bernevig}\ and\ \citenamefont {Zhang}(2006)}]{qsh1}%
  \BibitemOpen
  \bibfield  {author} {\bibinfo {author} {\bibfnamefont {B.~A.}\ \bibnamefont
  {Bernevig}}\ and\ \bibinfo {author} {\bibfnamefont {S.-C.}\ \bibnamefont
  {Zhang}},\ }\bibfield  {title} {\bibinfo {title} {Quantum spin hall effect},\
  }\href {https://doi.org/10.1103/PhysRevLett.96.106802} {\bibfield  {journal}
  {\bibinfo  {journal} {Phys. Rev. Lett.}\ }\textbf {\bibinfo {volume} {96}},\
  \bibinfo {pages} {106802} (\bibinfo {year} {2006})}\BibitemShut {NoStop}%
\bibitem [{\citenamefont {Moore}\ and\ \citenamefont {Balents}(2007)}]{tr1}%
  \BibitemOpen
  \bibfield  {author} {\bibinfo {author} {\bibfnamefont {J.~E.}\ \bibnamefont
  {Moore}}\ and\ \bibinfo {author} {\bibfnamefont {L.}~\bibnamefont
  {Balents}},\ }\bibfield  {title} {\bibinfo {title} {Topological invariants of
  time-reversal-invariant band structures},\ }\href
  {https://doi.org/10.1103/PhysRevB.75.121306} {\bibfield  {journal} {\bibinfo
  {journal} {Phys. Rev. B}\ }\textbf {\bibinfo {volume} {75}},\ \bibinfo
  {pages} {121306} (\bibinfo {year} {2007})}\BibitemShut {NoStop}%
\bibitem [{\citenamefont {Fu}\ and\ \citenamefont {Kane}(2006)}]{tr2}%
  \BibitemOpen
  \bibfield  {author} {\bibinfo {author} {\bibfnamefont {L.}~\bibnamefont
  {Fu}}\ and\ \bibinfo {author} {\bibfnamefont {C.~L.}\ \bibnamefont {Kane}},\
  }\bibfield  {title} {\bibinfo {title} {Time reversal polarization and a
  ${Z}_{2}$ adiabatic spin pump},\ }\href
  {https://doi.org/10.1103/PhysRevB.74.195312} {\bibfield  {journal} {\bibinfo
  {journal} {Phys. Rev. B}\ }\textbf {\bibinfo {volume} {74}},\ \bibinfo
  {pages} {195312} (\bibinfo {year} {2006})}\BibitemShut {NoStop}%
\bibitem [{\citenamefont {{Varjas}}(2022)}]{Daniel_pfaffian}%
  \BibitemOpen
  \bibfield  {author} {\bibinfo {author} {\bibfnamefont {D.}~\bibnamefont
  {{Varjas}}},\ }\bibfield  {title} {\bibinfo {title} {{Generalizations of the
  Pfaffian to non-antisymmetric matrices}},\ }\href
  {https://doi.org/10.48550/arXiv.2209.02578} {\bibfield  {journal} {\bibinfo
  {journal} {arXiv e-prints}\ ,\ \bibinfo {pages} {arXiv:2209.02578}} (\bibinfo
  {year} {2022})}\BibitemShut {NoStop}%
\bibitem [{\citenamefont {Cheng}\ \emph {et~al.}(2022)\citenamefont {Cheng},
  \citenamefont {Bomantara}, \citenamefont {Xue}, \citenamefont {Zhu},
  \citenamefont {Gong},\ and\ \citenamefont {Zhang}}]{pi_over_2_modes}%
  \BibitemOpen
  \bibfield  {author} {\bibinfo {author} {\bibfnamefont {Z.}~\bibnamefont
  {Cheng}}, \bibinfo {author} {\bibfnamefont {R.~W.}\ \bibnamefont
  {Bomantara}}, \bibinfo {author} {\bibfnamefont {H.}~\bibnamefont {Xue}},
  \bibinfo {author} {\bibfnamefont {W.}~\bibnamefont {Zhu}}, \bibinfo {author}
  {\bibfnamefont {J.}~\bibnamefont {Gong}},\ and\ \bibinfo {author}
  {\bibfnamefont {B.}~\bibnamefont {Zhang}},\ }\bibfield  {title} {\bibinfo
  {title} {Observation of $\ensuremath{\pi}/2$ modes in an acoustic floquet
  system},\ }\href {https://doi.org/10.1103/PhysRevLett.129.254301} {\bibfield
  {journal} {\bibinfo  {journal} {Phys. Rev. Lett.}\ }\textbf {\bibinfo
  {volume} {129}},\ \bibinfo {pages} {254301} (\bibinfo {year}
  {2022})}\BibitemShut {NoStop}%
\bibitem [{\citenamefont {Rudner}\ \emph {et~al.}(2013)\citenamefont {Rudner},
  \citenamefont {Lindner}, \citenamefont {Berg},\ and\ \citenamefont
  {Levin}}]{Rudner2013}%
  \BibitemOpen
  \bibfield  {author} {\bibinfo {author} {\bibfnamefont {M.~S.}\ \bibnamefont
  {Rudner}}, \bibinfo {author} {\bibfnamefont {N.~H.}\ \bibnamefont {Lindner}},
  \bibinfo {author} {\bibfnamefont {E.}~\bibnamefont {Berg}},\ and\ \bibinfo
  {author} {\bibfnamefont {M.}~\bibnamefont {Levin}},\ }\bibfield  {title}
  {\bibinfo {title} {Anomalous edge states and the bulk-edge correspondence for
  periodically driven two-dimensional systems},\ }\href
  {https://doi.org/10.1103/physrevx.3.031005} {\bibfield  {journal} {\bibinfo
  {journal} {Phys. Rev. X}\ }\textbf {\bibinfo {volume} {3}},\ \bibinfo {pages}
  {031005} (\bibinfo {year} {2013})}\BibitemShut {NoStop}%
\bibitem [{\citenamefont {Ozawa}\ \emph {et~al.}(2019)\citenamefont {Ozawa},
  \citenamefont {Price}, \citenamefont {Amo}, \citenamefont {Goldman},
  \citenamefont {Hafezi}, \citenamefont {Lu}, \citenamefont {Rechtsman},
  \citenamefont {Schuster}, \citenamefont {Simon}, \citenamefont {Zilberberg},\
  and\ \citenamefont {Carusotto}}]{Ozawa2019}%
  \BibitemOpen
  \bibfield  {author} {\bibinfo {author} {\bibfnamefont {T.}~\bibnamefont
  {Ozawa}}, \bibinfo {author} {\bibfnamefont {H.~M.}\ \bibnamefont {Price}},
  \bibinfo {author} {\bibfnamefont {A.}~\bibnamefont {Amo}}, \bibinfo {author}
  {\bibfnamefont {N.}~\bibnamefont {Goldman}}, \bibinfo {author} {\bibfnamefont
  {M.}~\bibnamefont {Hafezi}}, \bibinfo {author} {\bibfnamefont
  {L.}~\bibnamefont {Lu}}, \bibinfo {author} {\bibfnamefont {M.~C.}\
  \bibnamefont {Rechtsman}}, \bibinfo {author} {\bibfnamefont {D.}~\bibnamefont
  {Schuster}}, \bibinfo {author} {\bibfnamefont {J.}~\bibnamefont {Simon}},
  \bibinfo {author} {\bibfnamefont {O.}~\bibnamefont {Zilberberg}},\ and\
  \bibinfo {author} {\bibfnamefont {I.}~\bibnamefont {Carusotto}},\ }\bibfield
  {title} {\bibinfo {title} {Topological photonics},\ }\href
  {https://doi.org/10.1103/RevModPhys.91.015006} {\bibfield  {journal}
  {\bibinfo  {journal} {Rev. Mod. Phys.}\ }\textbf {\bibinfo {volume} {91}},\
  \bibinfo {pages} {015006} (\bibinfo {year} {2019})}\BibitemShut {NoStop}%
\bibitem [{\citenamefont {Hafezi}\ \emph
  {et~al.}(2011{\natexlab{a}})\citenamefont {Hafezi}, \citenamefont {Demler},
  \citenamefont {Lukin},\ and\ \citenamefont {Taylor}}]{Hafezi2011}%
  \BibitemOpen
  \bibfield  {author} {\bibinfo {author} {\bibfnamefont {M.}~\bibnamefont
  {Hafezi}}, \bibinfo {author} {\bibfnamefont {E.~A.}\ \bibnamefont {Demler}},
  \bibinfo {author} {\bibfnamefont {M.~D.}\ \bibnamefont {Lukin}},\ and\
  \bibinfo {author} {\bibfnamefont {J.~M.}\ \bibnamefont {Taylor}},\ }\bibfield
   {title} {\bibinfo {title} {Robust optical delay lines with topological
  protection},\ }\href {https://doi.org/10.1038/nphys2063} {\bibfield
  {journal} {\bibinfo  {journal} {Nat. Phys.}\ }\textbf {\bibinfo {volume}
  {7}},\ \bibinfo {pages} {907} (\bibinfo {year}
  {2011}{\natexlab{a}})}\BibitemShut {NoStop}%
\bibitem [{\citenamefont {Hafezi}\ \emph
  {et~al.}(2013{\natexlab{a}})\citenamefont {Hafezi}, \citenamefont {Mittal},
  \citenamefont {Fan}, \citenamefont {Migdall},\ and\ \citenamefont
  {Taylor}}]{Hafezi2013}%
  \BibitemOpen
  \bibfield  {author} {\bibinfo {author} {\bibfnamefont {M.}~\bibnamefont
  {Hafezi}}, \bibinfo {author} {\bibfnamefont {S.}~\bibnamefont {Mittal}},
  \bibinfo {author} {\bibfnamefont {J.}~\bibnamefont {Fan}}, \bibinfo {author}
  {\bibfnamefont {A.}~\bibnamefont {Migdall}},\ and\ \bibinfo {author}
  {\bibfnamefont {J.~M.}\ \bibnamefont {Taylor}},\ }\bibfield  {title}
  {\bibinfo {title} {Imaging topological edge states in silicon photonics},\
  }\href {https://doi.org/10.1038/nphoton.2013.274} {\bibfield  {journal}
  {\bibinfo  {journal} {Nat. Photonics}\ }\textbf {\bibinfo {volume} {7}},\
  \bibinfo {pages} {1001} (\bibinfo {year} {2013}{\natexlab{a}})}\BibitemShut
  {NoStop}%
\bibitem [{\citenamefont {Liang}\ and\ \citenamefont
  {Chong}(2013)}]{Liang2013}%
  \BibitemOpen
  \bibfield  {author} {\bibinfo {author} {\bibfnamefont {G.~Q.}\ \bibnamefont
  {Liang}}\ and\ \bibinfo {author} {\bibfnamefont {Y.~D.}\ \bibnamefont
  {Chong}},\ }\bibfield  {title} {\bibinfo {title} {Optical resonator analog of
  a two-dimensional topological insulator},\ }\href
  {https://doi.org/10.1103/PhysRevLett.110.203904} {\bibfield  {journal}
  {\bibinfo  {journal} {Phys. Rev. Lett.}\ }\textbf {\bibinfo {volume} {110}},\
  \bibinfo {pages} {203904} (\bibinfo {year} {2013})}\BibitemShut {NoStop}%
\bibitem [{\citenamefont {Afzal}\ \emph {et~al.}(2020)\citenamefont {Afzal},
  \citenamefont {Zimmerling}, \citenamefont {Ren}, \citenamefont {Perron},\
  and\ \citenamefont {Van}}]{exp2}%
  \BibitemOpen
  \bibfield  {author} {\bibinfo {author} {\bibfnamefont {S.}~\bibnamefont
  {Afzal}}, \bibinfo {author} {\bibfnamefont {T.~J.}\ \bibnamefont
  {Zimmerling}}, \bibinfo {author} {\bibfnamefont {Y.}~\bibnamefont {Ren}},
  \bibinfo {author} {\bibfnamefont {D.}~\bibnamefont {Perron}},\ and\ \bibinfo
  {author} {\bibfnamefont {V.}~\bibnamefont {Van}},\ }\bibfield  {title}
  {\bibinfo {title} {Realization of anomalous floquet insulators in strongly
  coupled nanophotonic lattices},\ }\href
  {https://doi.org/10.1103/PhysRevLett.124.253601} {\bibfield  {journal}
  {\bibinfo  {journal} {Phys. Rev. Lett.}\ }\textbf {\bibinfo {volume} {124}},\
  \bibinfo {pages} {253601} (\bibinfo {year} {2020})}\BibitemShut {NoStop}%
\bibitem [{\citenamefont {Wang}\ \emph {et~al.}(2017)\citenamefont {Wang},
  \citenamefont {Xiao}, \citenamefont {Liu}, \citenamefont {Zhu},\ and\
  \citenamefont {Chan}}]{exp3}%
  \BibitemOpen
  \bibfield  {author} {\bibinfo {author} {\bibfnamefont {Q.}~\bibnamefont
  {Wang}}, \bibinfo {author} {\bibfnamefont {M.}~\bibnamefont {Xiao}}, \bibinfo
  {author} {\bibfnamefont {H.}~\bibnamefont {Liu}}, \bibinfo {author}
  {\bibfnamefont {S.}~\bibnamefont {Zhu}},\ and\ \bibinfo {author}
  {\bibfnamefont {C.~T.}\ \bibnamefont {Chan}},\ }\bibfield  {title} {\bibinfo
  {title} {Optical interface states protected by synthetic weyl points},\
  }\href {https://doi.org/10.1103/PhysRevX.7.031032} {\bibfield  {journal}
  {\bibinfo  {journal} {Phys. Rev. X}\ }\textbf {\bibinfo {volume} {7}},\
  \bibinfo {pages} {031032} (\bibinfo {year} {2017})}\BibitemShut {NoStop}%
\bibitem [{\citenamefont {Hu}\ \emph {et~al.}(2015)\citenamefont {Hu},
  \citenamefont {Pillay}, \citenamefont {Wu}, \citenamefont {Pasek},
  \citenamefont {Shum},\ and\ \citenamefont {Chong}}]{Hu2015}%
  \BibitemOpen
  \bibfield  {author} {\bibinfo {author} {\bibfnamefont {W.}~\bibnamefont
  {Hu}}, \bibinfo {author} {\bibfnamefont {J.~C.}\ \bibnamefont {Pillay}},
  \bibinfo {author} {\bibfnamefont {K.}~\bibnamefont {Wu}}, \bibinfo {author}
  {\bibfnamefont {M.}~\bibnamefont {Pasek}}, \bibinfo {author} {\bibfnamefont
  {P.~P.}\ \bibnamefont {Shum}},\ and\ \bibinfo {author} {\bibfnamefont
  {Y.~D.}\ \bibnamefont {Chong}},\ }\bibfield  {title} {\bibinfo {title}
  {Measurement of a topological edge invariant in a microwave network},\ }\href
  {https://doi.org/10.1103/PhysRevX.5.011012} {\bibfield  {journal} {\bibinfo
  {journal} {Phys. Rev. X}\ }\textbf {\bibinfo {volume} {5}},\ \bibinfo {pages}
  {011012} (\bibinfo {year} {2015})}\BibitemShut {NoStop}%
\bibitem [{\citenamefont {Hafezi}\ \emph
  {et~al.}(2013{\natexlab{b}})\citenamefont {Hafezi}, \citenamefont {Mittal},
  \citenamefont {Fan}, \citenamefont {Migdall},\ and\ \citenamefont
  {Taylor}}]{Hafezi2013a}%
  \BibitemOpen
  \bibfield  {author} {\bibinfo {author} {\bibfnamefont {M.}~\bibnamefont
  {Hafezi}}, \bibinfo {author} {\bibfnamefont {S.}~\bibnamefont {Mittal}},
  \bibinfo {author} {\bibfnamefont {J.}~\bibnamefont {Fan}}, \bibinfo {author}
  {\bibfnamefont {A.}~\bibnamefont {Migdall}},\ and\ \bibinfo {author}
  {\bibfnamefont {J.~M.}\ \bibnamefont {Taylor}},\ }\bibfield  {title}
  {\bibinfo {title} {Imaging topological edge states in silicon photonics},\
  }\href {https://doi.org/10.1038/nphoton.2013.274} {\bibfield  {journal}
  {\bibinfo  {journal} {Nat. Photonics}\ }\textbf {\bibinfo {volume} {7}},\
  \bibinfo {pages} {1001} (\bibinfo {year} {2013}{\natexlab{b}})}\BibitemShut
  {NoStop}%
\bibitem [{\citenamefont {Gao}\ \emph {et~al.}(2018)\citenamefont {Gao},
  \citenamefont {Gao}, \citenamefont {Zhang}, \citenamefont {Luo},\ and\
  \citenamefont {Zhang}}]{Gao2018}%
  \BibitemOpen
  \bibfield  {author} {\bibinfo {author} {\bibfnamefont {Z.}~\bibnamefont
  {Gao}}, \bibinfo {author} {\bibfnamefont {F.}~\bibnamefont {Gao}}, \bibinfo
  {author} {\bibfnamefont {Y.}~\bibnamefont {Zhang}}, \bibinfo {author}
  {\bibfnamefont {Y.}~\bibnamefont {Luo}},\ and\ \bibinfo {author}
  {\bibfnamefont {B.}~\bibnamefont {Zhang}},\ }\bibfield  {title} {\bibinfo
  {title} {Flexible photonic topological insulator},\ }\href
  {https://doi.org/10.1002/adom.201800532} {\bibfield  {journal} {\bibinfo
  {journal} {Adv. Opt. Mater.}\ }\textbf {\bibinfo {volume} {6}},\ \bibinfo
  {pages} {1800532} (\bibinfo {year} {2018})}\BibitemShut {NoStop}%
\bibitem [{\citenamefont {Gao}\ \emph {et~al.}(2016)\citenamefont {Gao},
  \citenamefont {Gao}, \citenamefont {Shi}, \citenamefont {Yang}, \citenamefont
  {Lin}, \citenamefont {Xu}, \citenamefont {Joannopoulos}, \citenamefont
  {Soljačić}, \citenamefont {Chen}, \citenamefont {Lu}, \citenamefont
  {Chong},\ and\ \citenamefont {Zhang}}]{Gao2016}%
  \BibitemOpen
  \bibfield  {author} {\bibinfo {author} {\bibfnamefont {F.}~\bibnamefont
  {Gao}}, \bibinfo {author} {\bibfnamefont {Z.}~\bibnamefont {Gao}}, \bibinfo
  {author} {\bibfnamefont {X.}~\bibnamefont {Shi}}, \bibinfo {author}
  {\bibfnamefont {Z.}~\bibnamefont {Yang}}, \bibinfo {author} {\bibfnamefont
  {X.}~\bibnamefont {Lin}}, \bibinfo {author} {\bibfnamefont {H.}~\bibnamefont
  {Xu}}, \bibinfo {author} {\bibfnamefont {J.~D.}\ \bibnamefont
  {Joannopoulos}}, \bibinfo {author} {\bibfnamefont {M.}~\bibnamefont
  {Soljačić}}, \bibinfo {author} {\bibfnamefont {H.}~\bibnamefont {Chen}},
  \bibinfo {author} {\bibfnamefont {L.}~\bibnamefont {Lu}}, \bibinfo {author}
  {\bibfnamefont {Y.}~\bibnamefont {Chong}},\ and\ \bibinfo {author}
  {\bibfnamefont {B.}~\bibnamefont {Zhang}},\ }\bibfield  {title} {\bibinfo
  {title} {Probing topological protection using a designer surface plasmon
  structure},\ }\href {https://doi.org/10.1038/ncomms11619} {\bibfield
  {journal} {\bibinfo  {journal} {Nat. Commun.}\ }\textbf {\bibinfo {volume}
  {7}},\ \bibinfo {pages} {11619} (\bibinfo {year} {2016})}\BibitemShut
  {NoStop}%
\bibitem [{\citenamefont {Hafezi}\ \emph
  {et~al.}(2011{\natexlab{b}})\citenamefont {Hafezi}, \citenamefont {Demler},
  \citenamefont {Lukin},\ and\ \citenamefont {Taylor}}]{Hafezi2011a}%
  \BibitemOpen
  \bibfield  {author} {\bibinfo {author} {\bibfnamefont {M.}~\bibnamefont
  {Hafezi}}, \bibinfo {author} {\bibfnamefont {E.~A.}\ \bibnamefont {Demler}},
  \bibinfo {author} {\bibfnamefont {M.~D.}\ \bibnamefont {Lukin}},\ and\
  \bibinfo {author} {\bibfnamefont {J.~M.}\ \bibnamefont {Taylor}},\ }\bibfield
   {title} {\bibinfo {title} {Robust optical delay lines with topological
  protection},\ }\href {https://doi.org/10.1038/nphys2063} {\bibfield
  {journal} {\bibinfo  {journal} {Nat. Phys.}\ }\textbf {\bibinfo {volume}
  {7}},\ \bibinfo {pages} {907} (\bibinfo {year}
  {2011}{\natexlab{b}})}\BibitemShut {NoStop}%
\bibitem [{\citenamefont {El~Hassan}\ \emph {et~al.}(2019)\citenamefont
  {El~Hassan}, \citenamefont {Kunst}, \citenamefont {Moritz}, \citenamefont
  {Andler}, \citenamefont {Bergholtz},\ and\ \citenamefont
  {Bourennane}}]{ElHassan2019}%
  \BibitemOpen
  \bibfield  {author} {\bibinfo {author} {\bibfnamefont {A.}~\bibnamefont
  {El~Hassan}}, \bibinfo {author} {\bibfnamefont {F.~K.}\ \bibnamefont
  {Kunst}}, \bibinfo {author} {\bibfnamefont {A.}~\bibnamefont {Moritz}},
  \bibinfo {author} {\bibfnamefont {G.}~\bibnamefont {Andler}}, \bibinfo
  {author} {\bibfnamefont {E.~J.}\ \bibnamefont {Bergholtz}},\ and\ \bibinfo
  {author} {\bibfnamefont {M.}~\bibnamefont {Bourennane}},\ }\bibfield  {title}
  {\bibinfo {title} {Corner states of light in photonic waveguides},\ }\href
  {https://doi.org/10.1038/s41566-019-0519-y} {\bibfield  {journal} {\bibinfo
  {journal} {Nat. Photonics}\ }\textbf {\bibinfo {volume} {13}},\ \bibinfo
  {pages} {697} (\bibinfo {year} {2019})}\BibitemShut {NoStop}%
\bibitem [{\citenamefont {Fukui}\ \emph {et~al.}(2005)\citenamefont {Fukui},
  \citenamefont {Hatsugai},\ and\ \citenamefont {Suzuki}}]{Fukui_chern}%
  \BibitemOpen
  \bibfield  {author} {\bibinfo {author} {\bibfnamefont {T.}~\bibnamefont
  {Fukui}}, \bibinfo {author} {\bibfnamefont {Y.}~\bibnamefont {Hatsugai}},\
  and\ \bibinfo {author} {\bibfnamefont {H.}~\bibnamefont {Suzuki}},\
  }\bibfield  {title} {\bibinfo {title} {Chern numbers in discretized brillouin
  zone: Efficient method of computing (spin) hall conductances},\ }\href
  {https://doi.org/10.1143/JPSJ.74.1674} {\bibfield  {journal} {\bibinfo
  {journal} {J. Phys. Soc. Jpn.}\ }\textbf {\bibinfo {volume} {74}},\ \bibinfo
  {pages} {1674} (\bibinfo {year} {2005})}\BibitemShut {NoStop}%
\end{thebibliography}%

\end{document}